\documentclass[fleqn,usenatbib]{mnras}

\usepackage{newtxtext,newtxmath}

\usepackage[T1]{fontenc}

\DeclareRobustCommand{\VAN}[3]{#2}
\let\VANthebibliography\thebibliography
\def\thebibliography{\DeclareRobustCommand{\VAN}[3]{##3}\VANthebibliography}

\usepackage{graphicx}	
\usepackage{amsmath}	
\usepackage{caption}

\usepackage[export]{adjustbox}
\usepackage{subcaption}
\usepackage{graphicx}

\usepackage[utf8]{inputenc}

\title[Disc-wind-jet connection in quasars]{ Exploring the quasar disc-wind-jet connection with LoTSS and SDSS}

\author[Charlotte L. Jackson et al]{
Charlotte L. Jackson,$^{1}$\thanks{E-mail: charlotte.jackson@physics.ox.ac.uk}
James H. Matthews,$^{1}$
Imogen H. Whittam,$^{1,2}$
Matt J. Jarvis,$^{1,2}$
\newauthor ~Matthew J. Temple,$^{3}$
Amy L. Rankine,$^{4}$
and Paul C. Hewett $^{5}$
\\
$^{1}$Department of Physics, Astrophysics, University of Oxford, Denys Wilkinson Building, Keble Road, Oxford OX1 3RH, UK\\
$^{2}$Department of Physics and Astronomy, University of the Western Cape, Robert Sobukwe Road, 7535 Bellville, Cape Town, South Africa
\\
$^{3}$Centre for Extragalactic Astronomy, Department of Physics, Durham University, Durham DH1 3LE, UK\\
$^{4}$Institute for Astronomy, University of Edinburgh, Royal Observatory, Blackford Hill, Edinburgh EH9 3HJ, UK\\
$^{5}$Institute of Astronomy, University of Cambridge, Madingley Road, Cambridge CB3 0HA, UK
}
\date{Accepted  2026 January 7, Received 2026 January 7; in original form 2025 October 24}

\pubyear{\the\year{}}

\begin{document}
\label{firstpage}
\pagerange{\pageref{firstpage}--\pageref{lastpage}}
\maketitle

\begin{abstract}
We investigate the relationship between disc winds, radio jets, accretion rates and black hole masses of a sample of $\sim$100k quasars at z $\approx$ 2. Combining spectra from the 17th data release of the Sloan Digital Sky Survey (SDSS) with radio fluxes from the 2nd data release of the Low Frequency ARray (LOFAR) Two-Meter Sky Survey (LoTSS), we statistically characterise a radio loud and radio quiet population using a two-component Gaussian Mixture model, and perform population matching in black hole mass and Eddington fraction. We determine how the fraction of radio loud sources changes across this parameter space, finding that jets are most efficiently produced in quasars with either a very massive central black hole ($M_{\textrm{BH}} > 10^9 \textrm{M}_{\odot}$) or one that is rapidly accreting ($\lambda_{\textrm{Edd}}>0.3$). We also show that there are differences in the blueshift of the $\textrm{C}\,\textsc{iv}$ $\lambda$1549{\AA} line and the equivalent width of the $\textrm{He}\,\textsc{ii}$ $\lambda$1640{\AA} line in radio loud and radio quiet quasars that persist even after accounting for differences in the mass and accretion rate of the central black hole. Generally, we find an anti-correlation between the inferred presence of disc winds and jets, which we suggest is mediated by differences in the quasars' spectral energy distributions. The latter result is shown through the close coupling between tracers of wind kinematics and the ionising flux-- which holds for both radio loud and radio quiet sources, despite differences between their emission line properties-- and is hinted at by a different Baldwin effect in the two populations.

\end{abstract}




\section{Introduction}

Quasars are some of the most extreme objects in the universe, with luminosities reaching up to $\sim10^{41}$ W \citep[$\sim10^{48}$~erg~s$^{-1}$,][]{whatsinaname}, and central supermassive black holes with masses of up to $10^{10}$M$_{\odot}$ \citep{big}. Originally, quasars were discovered as anomalous sources with enormous luminosities across many frequencies, particularly in the radio, but with only a point-like optical counterpart, morphologically similar to a distant star \citep[][]{qso,kellermann}. However, despite radio emission playing a critical role in their original identification, we know now that many quasars are very faint or undetected in the radio \citep{sandage,kukula}, and amongst those that do show strong emission, many host impressive extended structure in the form of (ultra-)relativistic collimated polar radio jets. Exactly the mechanism or physical characteristic of a quasar that determines whether it will launch a jet is yet to be precisely established. Furthermore, whilst it is also thought that quasars often form another type of energetic outflow, wide-angled winds, how these co-exist or interact with formation of jets is not fully understood. The third crucial part of the puzzle is how these two types of outflow connect to the underlying accretion physics of the quasar. 

\subsection{Quasar Outflow Physics}

There are two dominant theoretical models for the launching of quasar jets: one that is black hole driven \citep[][BZ]{bz}, and another that derives energy from the accretion disc instead \citep[][BP]{BP}. The BZ mechanism involves the extraction of rotational energy from the central supermassive black hole via a strong poloidal magnetic field with open field lines threading the event horizon \citep{talbot2021blandford}. On the other hand, in the BP process, the energy fuelling a radio jet is instead extracted from the accretion disc by magnetic fields threading the disc, and the jet is launched centrifugally \citep{xie2012two}. Key differences between these two launching mechanisms include the region of influence of the magnetic field (in the BP process, the field must be distributed across a large radial extent of the disc, whereas in BZ, it is only important at the black hole's ergosphere; \citealt{ferreira:insu-03326978}); the spin dependence of the launching (the BZ mechanism strictly requires the black hole to be spinning, which is not the case for BP); and the efficiency of jet formation (it is thought that a BZ-like process would result in a significantly higher efficiency than a BP-like one; \citealt{nemmen2015efficiency}).

The BP mechanism can also be invoked as an explanation for the launching of accretion disc winds in quasars. In that picture, gas is ejected by centrifugal forces along the field lines threading the disc \citep{magwindsog,magwinds}. This is the magnetohydrodynamic (MHD) wind launching mechanism, and whilst there is good reason to expect that it may play an important role in outflows from the accretion disc, it is unclear currently what the observational signatures of the wind might look like, due to their dependence on the illusive magnetic field configuration in the disc \citep[e.g.][]{BP,contBH,fukumura2018variable}. 
 
Winds can also be launched via thermal pressure \citep{laha2021}. Thermally driven winds occur when X-rays produced in the inner accretion disc irradiate the outer disc, bringing material up to the Compton temperature, at which point it expands and becomes a thermally driven wind at radii where the local escape velocity is exceeded by the sound speed in the material \citep{begelman,woods,miz}.

The third commonly considered launching mechanism is radiative line driving. The  governing principle is that bound electrons in a plasma provide a greater-than-Thomson cross-section to incident light \citep{lucy,castor,owocki}. Crucially, for substantial momentum transfer to occur, there must be a delicate balance in the quasar spectral energy distribution (SED). A strong driving (near-) ultraviolet flux is needed to produce the required radiation pressure, alongside limited extreme ultraviolet (EUV) and X-ray production to prevent over-ionising the gas \citep{murraywinds,higginbottom,temple2023}.

\subsection{Observational Signatures of Outflows}

\label{sec:lines}
The tendency for quasars to emit prodigiously across the entire electromagnetic spectrum allows for rich analyses probing their physical processes \citep{elvis}. In the standard picture of a quasar \citep[i.e. that of][etc.]{osterbrock1993nature,introductiontoagn,beckmann2012active,whatsinaname}, emission of different wavelengths comes from different physical regions. In particular, ultraviolet and optical emission is thought to be produced primarily in the accretion disc, whereas X-rays are emitted from the innermost region of the disc, the so-called X-ray corona, and infrared emission is produced by the absorption and subsequent remission from a dusty torus. Observations at radio wavelengths can uncover sources of synchrotron emission, such as in jets, winds or shocks, and are not subject to absorption by dust due to their long wavelengths. 

Spectroscopy can provide great insight into a quasar's inner working through the observable properties of important emission and absorption lines. Outflows in the form of disc winds can be inferred by the presence of broad absorption lines or the blueshift of the $\textrm{C}\,\textsc{iv}$ $\lambda1549$\AA~emission line \citep{leighly,proga2000dynamics,richards2011}. $\textrm{C}\,\textsc{iv}$ is a collisionally excited line formed in the broad-line region (BLR) of a quasar, which is commonly observed to be blueshifted with respect to its expected laboratory wavelength. Whilst the exact cause of this blueshift is still under investigation \citep[e.g.][]{jameswinds}, a common assumption is that this is a signature of outflowing material, often interpreted as a disc wind. 

There is also a strong inverse correlation between the equivalent widths (EWs) of various ultraviolet emission lines, including $\textrm{C}\,\textsc{iv}$, and the ultraviolet continuum luminosity, known as the `Baldwin Effect' \citep{baldwin}. \cite{richards2011} showed that $\textrm{EW}_{\textrm{$\textrm{C}\,\textsc{iv}$}}$ also anti-correlates with the strength of the blueshift of the $\textrm{C}\,\textsc{iv}$ line. Unfortunately, the part of the SED responsible for ionising such lines, the EUV, is generally inaccessible due to absorption along the line of sight \citep{timlinthethird}. 
Fortunately, the $\textrm{He}\,\textsc{ii}$ $\lambda1640$\AA ~line is a simple, single-electron process, produced by the recombination of $\textrm{He}\,\textsc{iii}$ to $\textrm{He}\,\textsc{ii}$. It provides a relatively `clean' measure of the number of incident photons with an energy high enough to create $\textrm{He}\,\textsc{iii}$, which is 54.4eV and above, as the ratio of photons produced by recombination and subsequent line emission is of the order of unity. Therefore, the relative strength of the $\textrm{He}\,\textsc{ii}$ line, measured by its EW, is a good `photon counter' for the ionising EUV flux \citep{mathews,baskinheii,rankine2020bal,temple2023}. 

\label{sec:radio in RL/RQ}
Radio emission in quasars is associated with a wide range of processes, and displays the most dynamic range in luminosity of all electromagnetic wavelengths, spanning from essentially non-existent to $10^{30} $ W\,$\textrm{Hz}^{-1}$. For the very highest radio luminosities, it is clear that the dominant source of radiation has its origins in jets carrying relativistic electrons \citep{laor2008}, but for those sources at low and more intermediate radio luminosities, the primary source of the emission is less clear \citep[see][for a review]{panessa}. Previous studies have suggested that it could come from the X-ray corona \citep{laor2008}, relativistic particles accelerated in shocks within outflows \citep{zakamska,nims2015}, star formation \citep{condonsf, bonzini, padovani2016, rosario2020}, or weak/ frustrated jets \citep{bicknell, robbro}. These processes will presumably also play a role luminous radio quasars, just a sub-dominant one. 
Historically, sources have been divided into two populations: those that have very strong radio emission originating in a high-power jet, and those that have not. These are often referred to as `radio loud' (RL) and `radio quiet' (RQ) respectively. The existence of a true dichotomy in the distribution of radio properties of quasars has been the focus of intensive study for many years \citep[e.g.][]{condon1980,moderskidich, mcldich, ciradich, beaklinidich, MacFdich, zhangdich}, as has the question of what determines whether a quasar will be RL or RQ. So far, there have been no properties, outside of their radio emission, that have been identified as an accurate predictive tool of the radio loudness of an individual quasar.  

\subsection{Motivation and Aims of This Study}

There remain several key questions still to be answered in order to fully understand the outflow properties of quasars. Why do some quasars launch jets, but not others? How are quasar jets and disc winds launched, and what is the connection between these two modes of outflow? How do the outflow properties of a quasar relate to its accretion physics? Many authors have sought to better understand the answers to these questions; particularly relevant to our work are several recent studies focusing on emission line `demographics', building on the work of \cite{richards2011}. \cite{rankine2021} studied the connection between radio and outflow properties of quasars, finding differences in the $\textrm{C}\,\textsc{iv}$ and $\textrm{He}\,\textsc{ii}$ emission lines of RL and RQ objects, yet also sources with similar ultraviolet properties but strikingly different radio emission. \cite{temple2023} also found population-level differences in emission line properties, but this time focusing on the connection with accretion rate and black hole mass. In particular, they find a regime at high black hole mass and accretion rate that is associated with the strongest disc winds and softest SED, along with a prominent switch in ultraviolet emission line behaviour at an Eddington fraction of $\simeq0.1$. \cite{petley} suggest that different outflow processes dominate in quasars as a function of $\textrm{C}\,\textsc{iv}$ blueshift and $\textrm{EW}_{\textrm{$\textrm{C}\,\textsc{iv}$}}$, which they argue offers a proxy to Eddington-scaled accretion rate. They point to high accretion rates corresponding primarily to disc wind-caused radio emission, and lower accretion rates associated with emission dominated by radio jets. These three specific works fit within a broader landscape of literature exploring how the radio emission in quasars depends on, for example, $\textrm{C}\,\textsc{iv}$ emission line properties \citep{richards2011,richards2021,kratzer2015,Calistro2024}, colour/dust extinction \citep{white2003,klindt2019,rosario2020,fawcett2020,fawcett2023} and the presence or absence of broad absorption lines \citep{stocke1992,hewett2003,morabito2019,petley2022}.

In our work, we connect these previous studies and further investigate the relationship between radio emission, accretion disc outflows, and fundamental properties of quasars, including Eddington fraction and black hole mass. The aim of this paper is to compare observational signatures of disc winds, radio jets, and the accretion disc in quasars. We use a combination of rest-frame ultraviolet spectroscopy with deep low frequency radio observations across $\sim$100 000 optically identified quasars, with a redshift range of \(1.5<z<2.65\), that gives coverage of the $\textrm{C}\,\textsc{iv}$, $\textrm{He}\,\textsc{ii}$ and $\textrm{Mg}\,\textsc{ii}$ emission lines within the Sloan Digital Sky Survey (SDSS) observational window. 
In Section \ref{sec:obs data}, we present the observational data and detail how our radio loud sample is defined. Our key results are presented in Section \ref{sec:results}, and we discuss their physical consequences in Section \ref{sec:discuss}. Throughout the paper, we give wavelengths in vacuum in units of \AA ngstrom, black hole masses in $\textrm{M}_{\odot}$, and luminosities in W\,Hz$^{-1}$. We assume a flat $\Lambda$CDM cosmology with $H_0 = 71~ 
 \textrm{kms}^{-1}~\textrm{Mpc}^{-1}$. Energies, frequencies and wavelengths are given in the rest-frame.

\section{Observational Data and Method}
\label{sec:obs data}
\begin{figure*}
    \centering
    \includegraphics[width=\linewidth]{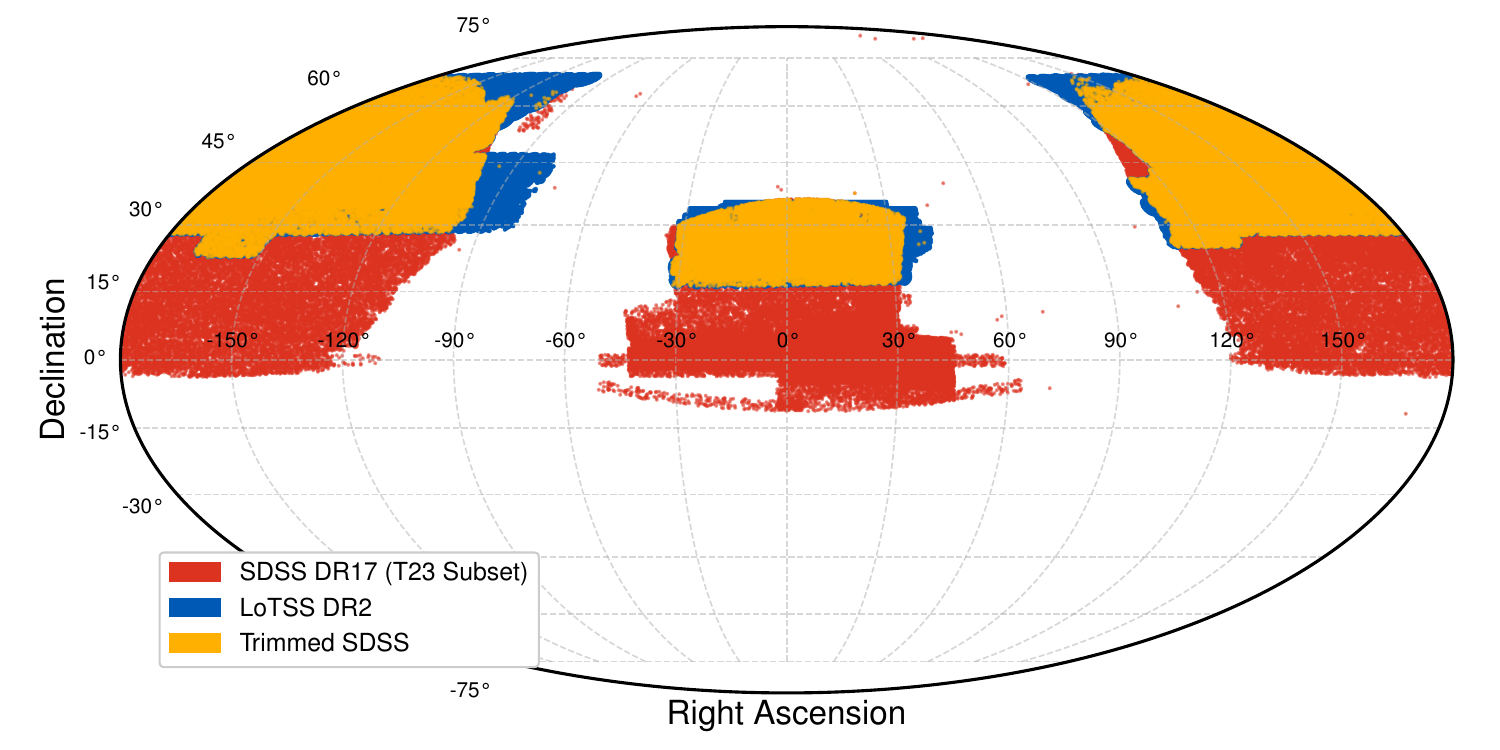}
    \caption{Sky coverage of the surveys used in this work. The Sloan Digital Sky Survey DR17 subset constructed by \protect\cite{temple2023} (red), LOFAR Two-Meter Sky Survey DR2 optical identifications catalogue (blue), and the resulting overlapping sample (yellow).}
    \label{fig:sky}
\end{figure*}

\subsection{Observational Data}
\label{sec:obs ii}

We use the subsample of quasars from the SDSS DR17, described by \cite{sdssdr16} and \cite{sdssdr17}, that was constructed by \cite{temple2023}. In their work, they carried out spectral reconstructions, based on the Mean-Field Independent Component Analysis performed by \cite{rankine2020bal}, and extracted spectroscopic redshifts and the emission line properties of $\textrm{C}\,\textsc{iv}$, $\textrm{He}\,\textsc{ii}$, and $\textrm{Mg}\,\textsc{ii}$. 186\,303 sources were both in the redshift range of \(1.5<z<2.65\) and met their spectral quality criteria. See \cite{temple2023} and \cite{rankine2020bal} for full details on the sample processing. 

Radio data for our sample are taken from the LOFAR Two-meter Sky Survey (LoTSS) DR2, described by \cite{shimwell2022LOFAR}. LoTSS is a wide-area radio survey operating at 144 MHz, with a sensitivity of 71 $\mu$Jy $\textrm{beam}^{-1}$. The second data release provides a catalogue of nearly 4.5 million sources at 6 arcsec resolution, covering 27 per cent of the northern sky, observing nearly an order of magnitude more sources than DR1. There is also excellent multi-wavelength coverage across the observing area of LoTSS, including significant overlap with SDSS DR17.

To obtain a combined radio and rest-frame ultraviolet catalogue, we cross-match the SDSS sample from \cite{temple2023} with the value-added LoTSS DR2 source catalogue described by \cite{hardcastle}, using the positions of the established optical counterparts of the radio sources. We first identified the subset of the quasar catalogue lying within the sky coverage of LoTSS DR2, an exercise which reduces the sample from 186 303 sources to 108 167. The sky coverage of the two datasets and the subsequently trimmed SDSS catalogue can be seen in Fig. \ref{fig:sky}. 
We then cross-matched the two catalogues with a 1 arcsec error radius. The cross-matching resulted in 15\,255 radio detected quasars, with a radio detection fraction of 15 per cent, which is similar to the results of \cite{rankine2021}, who cross-matched the LoTSS DR1 sample \citep{dr1} with SDSS DR14, and found an overall detection fraction of 16 per cent. The luminosities of the quasars in our sample spans a range of $L_{\textrm{144}} \approx 10^{24- 29}$ W\,Hz$^{-1}$.

For the other 92\,912 objects below the 5$\sigma$ LoTSS peak detection limit, we extract a radio flux density using forced photometry. In previous studies, the survey detection limit has been used to produce an upper bound on the radio luminosity of undetected sources, allowing them to be used in the radio quiet population, for example in \cite{rankine2021}. In this work, we elect to retrieve a flux density measurement directly from the LoTSS image to gain insight into the radio emission occurring below the catalogue threshold, and to confirm that these quasars are definitely radio quiet by all definitions. 
We downloaded 30 arcsec radio cutouts from the LOFAR API service\footnote{information available at \url{https://LOFAR-surveys.org/cutout_api_details.html}}, centred on their SDSS coordinates. The flux at the central pixel is extracted, and the median flux of 500 randomly selected pixels in the cutout is calculated as the background noise. The difference between the two is then taken as their radio flux density, which allows us to add the radio measurements of an additional 70\,292 sources to our sample. The central flux density of our undetected sources is taken as a good approximation for total flux as it is assumed that, if undetected in the LoTSS DR2 catalogue, the quasars are very likely to be unresolved point sources in the radio image, and as the radio image pixels are calibrated in flux/beam, it is ensured that, for unresolved sources, the peak flux is necessarily equal to the total flux. We verified that the pixel fluxes provide a good estimate of the total flux for the faintest sources in the detected sample. 

The black hole masses of the quasars in our sample are taken from the \cite{temple2023} catalogue, which used the single epoch virial estimator described by \cite{mattmasses} and later \cite{vestergaard}, as seen in equation \ref{eq: black hole mass}, 
\begin{equation} \label{eq: black hole mass}
   M_{\textrm{BH}} = 10^{6.86}\left(\frac{\textrm{FWHM}_{\textrm{$\textrm{Mg}\,\textsc{ii}$}}}{\textrm{1000 kms}^{-1}}\right)^2 \left(\frac{L_{3000}}{10^{44}\textrm{ergs}^{-1}}\right)^{0.5}\textrm{M}_{\odot}\, ,
\end{equation}
where $\textrm{FWHM}_{\textrm{$\textrm{Mg}\,\textsc{ii}$}}$ is the full-width at half maximum of the $\textrm{Mg}\,\textsc{ii}$ line, and $L_{3000}$ is the rest-frame monochromatic continuum luminosity at 3000\AA. $\textrm{Mg}\,\textsc{ii}$ is a strong, broad, low-ionisation line; the FWHM of which is thought to trace the viralised bulk motion of the emitting BLR. It is also known from reverberation mapping measurements \citep[e.g.][]{wandel_central_1999, kaspi_reverberation_2000} that there is a generally constant relationship between the physical size of the BLR and the observed UV luminosity, which \citealt{mattmasses} demonstrated is particularly tight for $L_{3000\AA}$. As such, the combination of these two properties can offer an estimation of the black hole mass of a quasar, through virial arguments. To ensure self-consistency, we therefore use $L_{3000}$ throughout the paper, including in the calculation of bolometric luminosity, $L_{\textrm{bol}} = 5.15 \times L_{3000}$, assuming a constant bolometric correction \citep{richards2006,shen2011}.  $L_{3000}$ has the advantage over monochromatic line luminosities (e.g. [$\textrm{O}\,\textsc{iii}$], $\textrm{H}\,\beta$) of offering a more direct view of a quasar's thermal luminosity \citep[i.e. the 'big blue bump',][]{punsly_calibrating_2011}. However, there are still multiple assumptions embedded in the use of constant bolometric correction, an overall discussion of which is given in section 5.1.1 of \cite{temple2023}. The Eddington fraction is then calculated as the ratio of a source's bolometric luminosity to its Eddington luminosity, \(L_{\textrm{bol}}/{L_{\textrm{Edd}}}\), herein \(\lambda_{\textrm{Edd}}\), given by:
\begin{equation} \label{eq: Ledd}
    \lambda_{\textrm{Edd}} = \frac{L_{\textrm{bol}}}{L_{\textrm{Edd}}} = \frac{5.15 \times L_{3000}}{1.26 \times 10^{38} \left(\frac{M_{\textrm{BH}}}{\textrm{M}_{\odot}}\right) \textrm{ergs}{^{-1}}} \, .
\end{equation}\\
The monochromatic radio luminosity at 144 MHz are calculated as
\begin{equation} \label{eq: lrad}
L_{\textrm{144}} = 4\pi d_L^2 S_{144\textrm{MHz}}(1+z)^{-\alpha-1}~ \textrm{W\,Hz$^{-1}$} \, ,
\end{equation}
where $d_L$ is the luminosity distance, $S_{144\textrm{MHz}}$ is the flux at 144 MHz, $z$ is the redshift, and $\alpha$ is the spectral index. Throughout the paper, we use $F_{\nu} \propto \nu^{\alpha}$, where $\alpha$ is assumed to be $-0.7$ \citep{kukula}. 
\begin{figure*}
    \includegraphics[width=\textwidth]{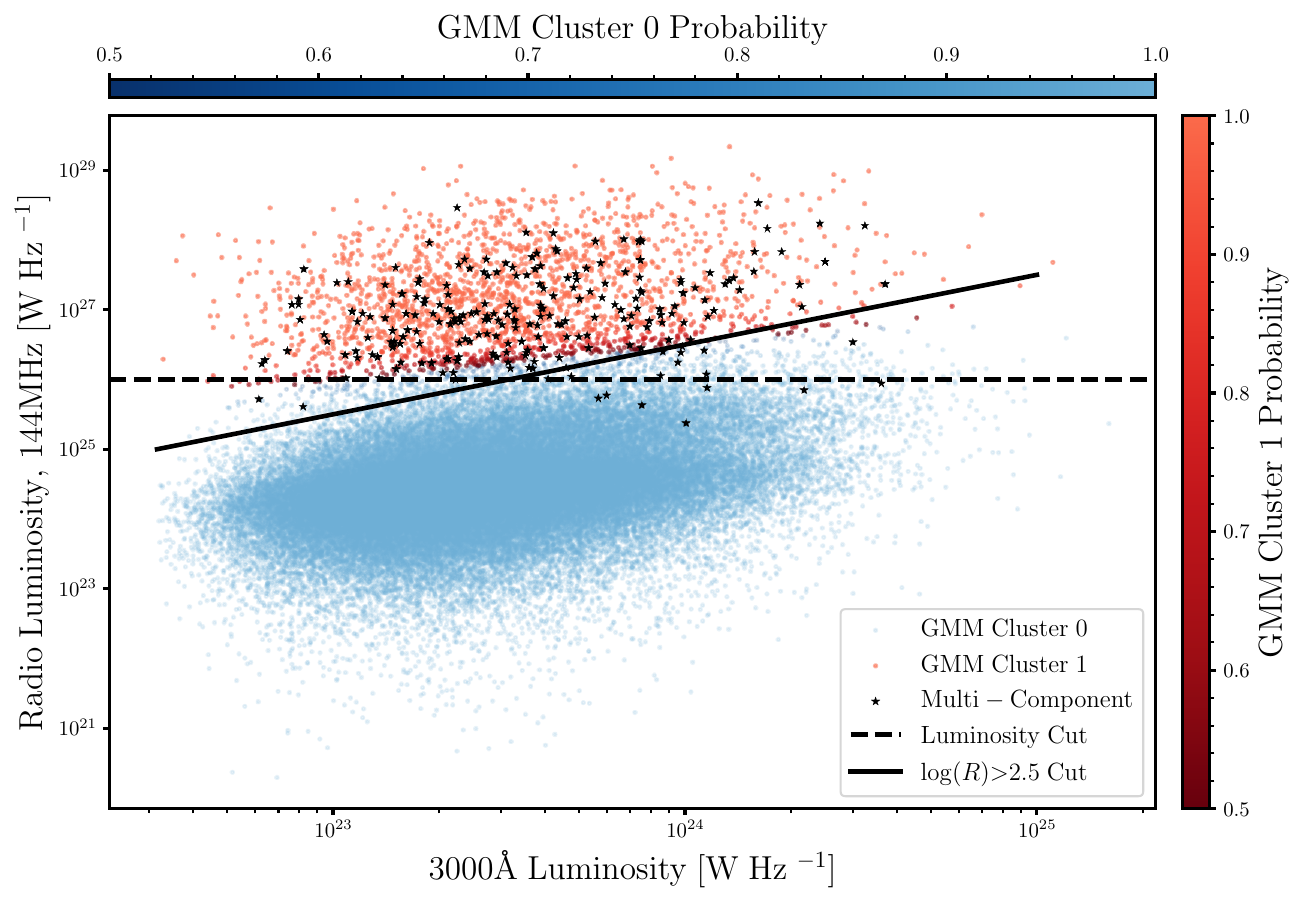}
    \caption{Distribution of quasars in $L_{3000}$ and 144~MHz radio luminosity. The colour of the points corresponds to the cluster to which the source has been assigned by the two-component GMM, with Cluster 1 (red) defining radio-loud and Cluster 2 (blue) defining radio-quiet sources. The shading of the points indicates the probability of the source being correctly assigned to its cluster, with darker colours showing less certainty. The solid black line shows the classical radio loudness cut at log$_{10}(R)>2.5$, and the dashed line indicates the $10^{26}$ W\,Hz$^{-1}$ radio luminosity cut. The stars show sources with multiple ($>1$) associated components in the LoTSS DR2 value-added catalogue.}
    \label{fig:rcut}
\end{figure*}

\subsection{Defining a `Jetted' Population}
\label{sec:how RL}
In the absence of high resolution very long baseline interferometry imaging and subsequent morphological classification for each of the $\sim \textrm{100k}$ objects, the identification of radio jets becomes a task of inference rather than direct observation. 

Standard practice for some time has been to divide quasars into two populations based on the `radio loudness' of a source, as defined by the ratio \( R = \frac{L_{\textrm{rad}}}{L_{\textrm{opt}}}\),
which normalises the radio luminosity by the optical luminosity. The normalisation allows for the identification of sources dominated by radio emission, rather than being biased towards those that are simply intrinsically brighter at all wavelengths. However, there are associated pitfalls with the method. First, there is the decision of where to place a cut-off between `radio loud' and `radio quiet' populations in a distribution which, with the development of deeper and more sensitive radio surveys such as LoTSS, has become increasingly continuous -- see. Fig. \ref{fig:rcut}. Too low a division, and non-jetted sources with strong radio emission originating from other processes, such as star formation, will contaminate the sample. Too high, and it risks missing weaker jets, particularly where there is relatively strong optical emission. 
Secondly, there is the interdependency between the ratio and other values that are calculated from the optical luminosity, such as black hole mass and Eddington fraction. As a result, it is more difficult for a source to meet the threshold if it has a higher black hole mass, based on its higher optical luminosity at fixed Eddington fraction, see equation \ref{eq: black hole mass}. 

One option to lift the optical luminosity dependency is to use a purely radio luminosity-based cut, which is more reminiscent of early work \citep[e.g.][]{miller1990bimodal,bestrlfrac}. Whilst the approach may result in an over-representation of sources that are simply the most intrinsically bright across all wavelengths, it should equally feature some of the more intermediate luminosity jets. However, this method is not used often in modern studies as it does not account for any correlations between radio and optical luminosities \citep[e.g.][]{balokovic}, and relies on a relatively arbitrary placement for the cut. 

Ideally, one would define a method to classify a jetted population based on the distribution of sources in the optical and radio luminosity plane, looking for a general minimum across the 2D distribution, which may not be linear. Previous works have sought to address this challenge, including \cite{bohan1,bohan2}, who used a parametric model describing the AGN and star formation contributions to a population's radio emission in an effort to statistically disentangle them. Another option, and the one we choose to use in this study, is to use a Gaussian Mixture Model (GMM), that fits Gaussian components to the overall distribution of sources, and defines `clusters'. As such, the statistical significance of the populations can be ascertained, and probabilities for each source being in each population quantified -- moving away from a blunt classifier based on a chosen cut location. 

For the purposes of this study, we choose to define a `radio loud' and `radio quiet' population using each of these three methods, and present the results of these classifications in Fig. \ref{fig:rcut}. We choose to use a GMM with two components, which is motivated by the distribution of the data and which has the benefit of producing RL and RQ analogues that are directly comparable to the use of a radio loudness or luminosity cut. However, many more Gaussians could in theory be fitted to the distribution, for example defining a `radio intermediate' cluster. Throughout the paper, we will discuss the impact of these different definitions.

We calculate the classical radio loudness parameter, $R$, of each source using the radio luminosity derived from a combination of the catalogue and measured radio flux densities, along with the spectroscopic redshifts of the quasars. We use the rest-frame monochromatic continuum luminosity at 3000{\AA} for $L_{\textrm{opt}} $, derived from \cite{sed}'s quasar SED model. The choice of radio loudness cut-off location depends largely on the scientific question aiming to be answered. In this work, we seek to isolate a population of jetted quasars, so choose a relatively strict cut of \(\textrm{log}_{10}(R)=2.5\), which corresponds to a minimum in the distribution of the $L_{144}$ and $L_{3000}$ (see Fig. \ref{fig:rcut}). It also depends heavily on the radio frequency used for observations, and therefore a spectral index assumption. The canonical threshold is \(\textrm{log}_{10}(R)=1\) at 5 GHz \citep{kellermann}, and at LOFAR frequencies, often a value of \(\rm log_{10}(\textit{R})>2\) is used. However, it should be noted that \cite{rankine2021} uses the SDSS i-band magnitude, which is not the same as our $L_{\textrm{3000}} $ measurement.  Our choice of cut undoubtedly means that jetted sources `leak' into the RQ population, but it is reasonable to assume that the RL subsample is relatively clean.  With the cut applied, out of the 105 784 sources in our sample, 2383 are classified as RL (a global proportion of 2.20 per cent), and the other 105 784 as RQ (see Table \ref{tab:cut stats}).\\

We also define RL and RQ populations based on the radio luminosity of our sources, removing any dependence on the emission from the accretion disc. We use $L_{\textrm{144}} > 10^{26} ~\textrm{W\,Hz$^{-1}$}$, above which, the radio emission is unlikely being produced by star formation alone. According to the relationship found by \cite{calistro2017LOFAR}, 
\begin{equation}{\label{eq:sfr}}
L_{\textrm{SFR}, \textrm{144MHz}} = \textrm{SFR} [\textrm{M}_{\odot}/ \textrm{yr}]\times{1.455 \times 10^{24} \times 10^{-q(z)}} \textrm{W\,Hz$^{-1}$}
\end{equation}
where $q(z) = 1.77(1+z)^{-0.22}$, a radio luminosity of $L_{\textrm{144}} > 10^{26} ~\textrm{W\,Hz$^{-1}$}$ would correspond to a rate on the order of 3000 $\textrm{M}_{\odot}$yr$^{-1}$, which is an order of magnitude greater than found in studies such as \cite{floydsfr} and \cite{harrissfr}. The criteria yields a global proportion of 2.38 per cent RL sources, the highest of the three methods. The trend of increasing $L_{\textrm{144}}$ with $L_{\textrm{3000}}$ means that sources are more likely to meet the threshold if they have a greater bolometric luminosity.  

Finally, we employ a GMM from \textsc{Scikit Learn} \citep{sklearn} with two components and a tied covariance matrix, for the sources in our sample with positive radio flux densities. Those with negative fluxes are automatically assigned to the RQ population, and are naturally included in the RQ classification of the other methods. Encouragingly, the two Gaussian components that fit to the $L_\textrm{rad}$-$L_{3000}$ distribution define clusters that are similar to the traditional radio loudness classifications. Almost all objects that are defined by the GMM as being RL meet the  $L_{\textrm{144}} > 10^{26} ~\textrm{W\,Hz$^{-1}$}$ criteria, and 1.86 per cent of all sources are classified as RL.  Furthermore, we are also able to assign a probability of each source being in the associated clusters. We use these as weights instead of a binary classification in the following analyses, as detailed in the results. \\
\begin{table}
    \centering
    \caption{The number of sources classified as radio loud and radio quiet for each of the three characterisation methods. The identification column shows how many sources with multiple associated components in the radio catalogue have been classified as radio loud per method, which we use as a test for success. Particularly, the GMM identifies the same number of multi-component objects for the fewest overall RL classifications, suggesting that it may be the most efficient.}
    \begin{tabular}{ccccc}
        \hline
        Cut Used & RL & RQ & Identified \\
        \hline\hline
         \(\rm log_{10}(\textit{R})>2.5\)  &2383& 105 784 & 221/231 \\
        \hline
        $L_{\textrm{144}} > 10^{26} \textrm{W\,Hz$^{-1}$}$ & 2577 &105 590 & 227/231\\
        \hline
        GMM & 2014 & 106 153 & 221/231\\
        \hline
    \end{tabular}
    
    \label{tab:cut stats}
\end{table}

In the absence of an accessible `ground truth' about which sources have jet activity, we use the LoTSS DR2 component catalogue produced by \cite{hardcastle} to identify sources with multiple associated components, which is an indicator of extended structure, and test whether the three classification methods `recover' these sources. However, extended structure is not guaranteed to be {\em jetted} structure, and there are most certainly sources with jets missing from the classification.  Therefore, we both can neither take a classification of `multi-component' as the smoking gun of a jet, nor the lack of classification as confirmation there is no jet present; nevertheless, the exercise acts as a reasonable sanity check. We find 231 objects present in our sample that have multiple components, which are overplotted on Fig. \ref{fig:rcut}. We present in Table \ref{tab:cut stats} the recovery rate for each classification method, which is consistent across all three and shows that, in general, each method is successful at finding jetted sources. There are 4 sources with multiple components that are not recovered at all, but these do not show convincing signs of jet activity in their radio image cutouts. The GMM classifies the least number of quasars as RL, but still recovers the same number of multi-component sources at the traditional radio loudness cut, suggesting that the method may produce the least contamination. Of objects that meet all three radio loudness criteria, 11.4 per cent are identified as having multiple components, whereas it is only 0.016 per cent for sources meeting none. 1.75 per cent of objects that meet the radio luminosity criteria but not the classical radio loudness cut (i.e. are in the high $L_{3000}$ `wedge' section of Fig. \ref{fig:rcut}) have multiple components, and 0.47 per cent of those meeting the classical radio loudness but not radio luminosity cuts (i.e. in the lower $L_{3000}$ wedge) have multiple components. \\

Throughout the rest of the paper, we present the results from the GMM-defined populations as standard, as the approach is data-driven and agnostic to the expected populations, and most efficiently retrieves the multi-component sources. Using the GMM also allows us to assign classification likelihoods to sources, allowing us to fold uncertainty in classification into our analyses. The results from the other two cuts are presented in Appendix \ref{Ap:diff cut}. In summary, the overall conclusions do not change between classification method, although the connection between $L_{3000}$ and black hole mass results in the weakening of some trends when making a cut in $L_{144}$ alone. All sources that are classified as RL by any method are detected in the original LoTSS DR2 catalogue, and thus do not rely on fluxes derived from forced photometry. 

\begin{figure*}
    \centering
    \includegraphics[width=\columnwidth]{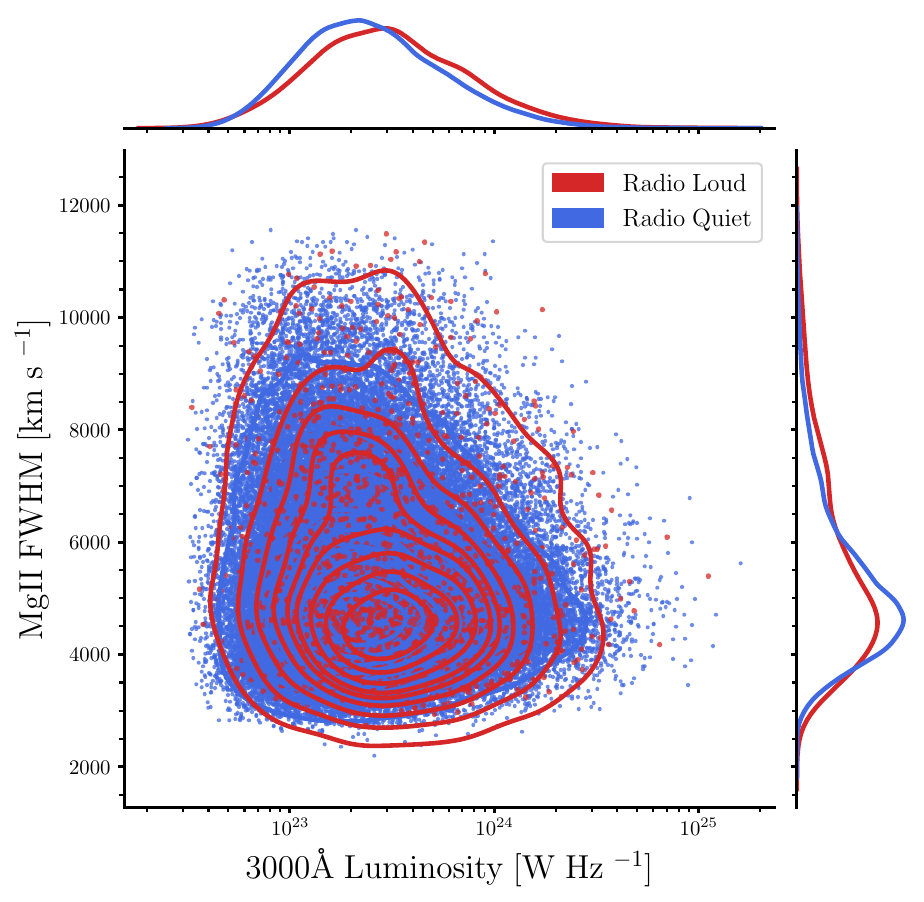}
    \includegraphics[width=\columnwidth]{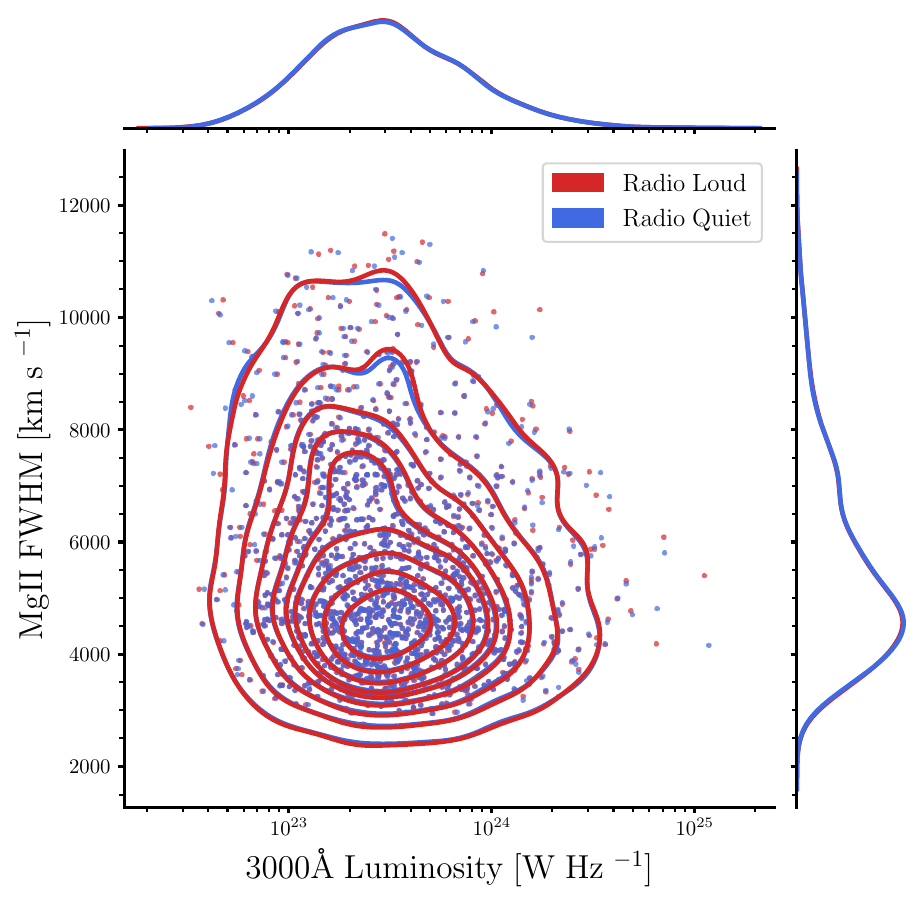}
    \caption{Distributions of radio loud (red points/contours) and radio quiet (blue points/contours) quasars in $L_{3000}$ and $\textrm{FWHM}_{\textrm{$\textrm{Mg}\,\textsc{ii}$}}$ space, for the entire sample (left) and then after NearestNeighbour matching in this space (right). The RL and RQ populations are as defined by the GMM. The matching procedure has removed any differences in these distributions.}
    \label{fig:pre-match}
\end{figure*} 

\subsection{Population Matching}

One of the principal aims of this paper is to understand how the presence of a jet is linked to other outflow and accretion properties in quasars. One test is the comparison of the emission line characteristics of our RL/RQ populations, such as $\textrm{C}\,\textsc{iv}$ blueshift, $\textrm{EW}_{\textrm{$\textrm{C}\,\textsc{iv}$}}$ and $\textrm{EW}_{\textrm{$\textrm{He}\,\textsc{ii}$}}$. In the LoTSS DR1 sample, \cite{rankine2021} show that RL quasars systemically display lower $\textrm{C}\,\textsc{iv}$ blueshifts than RQ ones. However, \cite{richards2011} demonstrated that $\textrm{C}\,\textsc{iv}$ blueshift increases with an increasing luminosity. As RL sources will inherently have a lower optical luminosity at a given radio luminosity than RQ, it is therefore possible that the weaker blueshifts are an artifact of the underlying trend. Alternatively, previous works have found evidence to suggest that RL quasars are generally associated with more massive central black holes \citep[e.g.][]{metcalf, laor2000, McJarvisLumMassRel, chakraborty}. It has also been suggested \citep[for example, by][]{progastini,petley} that there is a link between accretion rate and quasars jets, specifically, that a lower accretion rate is expected in the regime where jets dominate over winds. A connection has also been found between accretion rate and $\textrm{C}\,\textsc{iv}$ blueshift by \cite{wang2011} and \cite{sulentic2017}, which was extended to two dimensions by \cite{temple2023}, showing that $\textrm{C}\,\textsc{iv}$ blueshift is endemic only for the most massive, highly accreting black holes. For EWs, there is the well-known Baldwin Effect that leads one to expect weaker $\textrm{C}\,\textsc{iv}$ and $\textrm{He}\,\textsc{ii}$ in the sources more luminous in the ultraviolet, which is seen in \cite{temple2023} to result in deficiencies in these quantities for the most massive, highly accreting black holes also. \cite{rankine2021} equally shows that there is stronger overall $\textrm{He}\,\textsc{ii}$ emission in RL quasars.

Overall, it is clear that there is a complex interplay of characteristics that could lead to perceived correlations between emission line and radio properties, primarily related to differences in black hole mass and accretion rate distributions, or, equivalently, luminosity. Any relationships in this space need to be carefully examined to ensure they are not artificially induced, nor driven by underlying correlations in properties unrelated to the question at hand. \cite{stepneymatch} use population matching for their sample of SDSS DR16 quasars at $1.5<z<4$ to explore the evolution of their ultraviolet emission line properties with redshift. They examine the dependence on ultraviolet luminosity, or both black hole mass and Eddington-scaled accretion rate, and find that differences in $\textrm{C}\,\textsc{iv}$ blueshifts between different redshift bins can be fully explained by different distributions in luminosities between the populations, as these trends vanish once the matching has taken place. As such, for the populations in our study, if there are observable trends or differences in emission line properties that persist after the matching, we can be sure they are not driven by differences in the black hole masses, Eddington fraction, or $L_{3000}$ of RL and RQ quasars.  

We therefore match our RL and RQ populations in $L_{\textrm{3000}} $ and $\textrm{FWHM}_{\textrm{$\textrm{Mg}\,\textsc{ii}$}}$ space, as these are the fundamental observed quantities that are used to derive the Eddington fraction and black hole mass of the sources. The matching serves equally well to match them in those quantities, too.  We perform a two-dimensional (2D) K-nearest neighbours matching procedure using the {\tt NearestNeighbours} function from \textsc{Scikit Learn} \citep{scikit-learn}, with $k=1$, i.e. a single RQ neighbour returned. None of the RL sources are removed, and some of them share a nearest neighbour RQ source, resulting in 1976 RQ and 2014 RL quasars in the matched sample. In Fig. \ref{fig:pre-match}, we show the distribution of RL and RQ sources (as defined by the GMM) in $L_{\textrm{3000}} $  and $\textrm{FWHM}_{\textrm{$\textrm{Mg}\,\textsc{ii}$}}$ before and after matching. To verify the quality of the match, we carry out a 2D Kolmogorov--Smirnov (KS) test from \textsc{Scipy} \citep{scipy}. The original distribution has a $p$-value of $10^{-19}$, where anything $<0.05$ allows us to reject the null hypothesis that the distributions are drawn from the same parent population at the $95$ per cent confidence level. After matching, the $p$-value is 1.0 indicating statistically similar 2D distributions. The redshift distributions of our RL and RQ populations are virtually indistinguishable both before and after matching.

\section{Results}
\label{sec:results}

\subsection{Emission Line Properties of RL and RQ Sources}
\label{sec:emission lines}
\begin{figure*}
    \centering
    \includegraphics[width=\columnwidth]{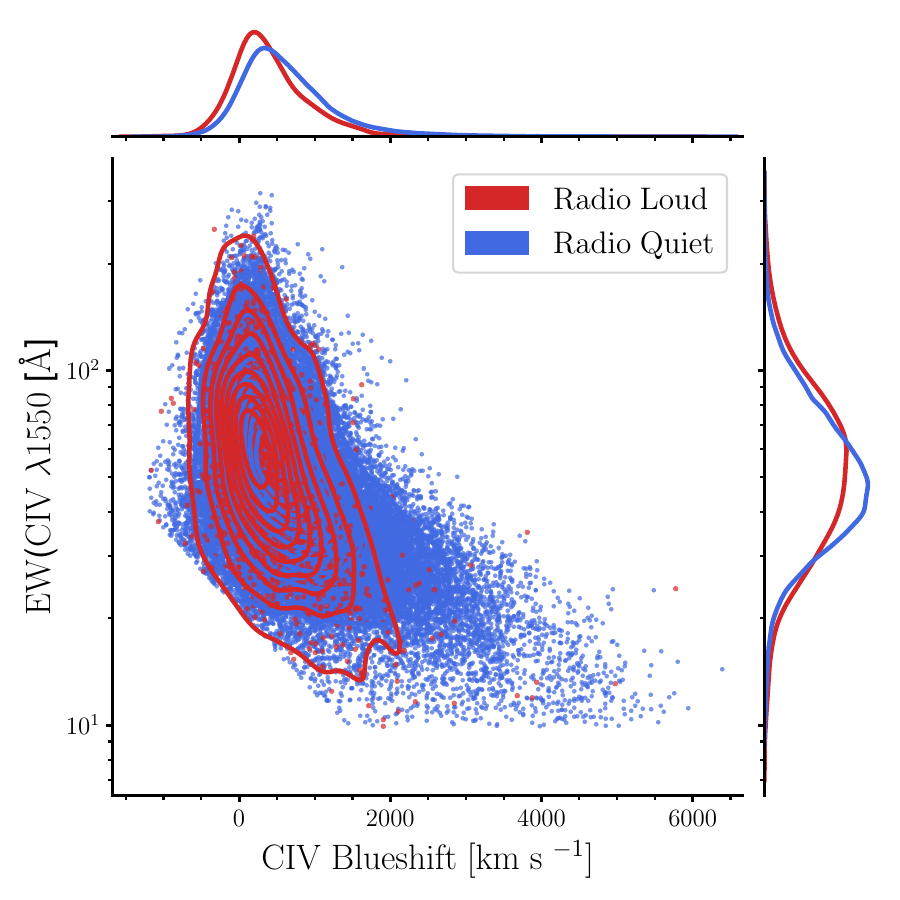}
    \includegraphics[width=\columnwidth]{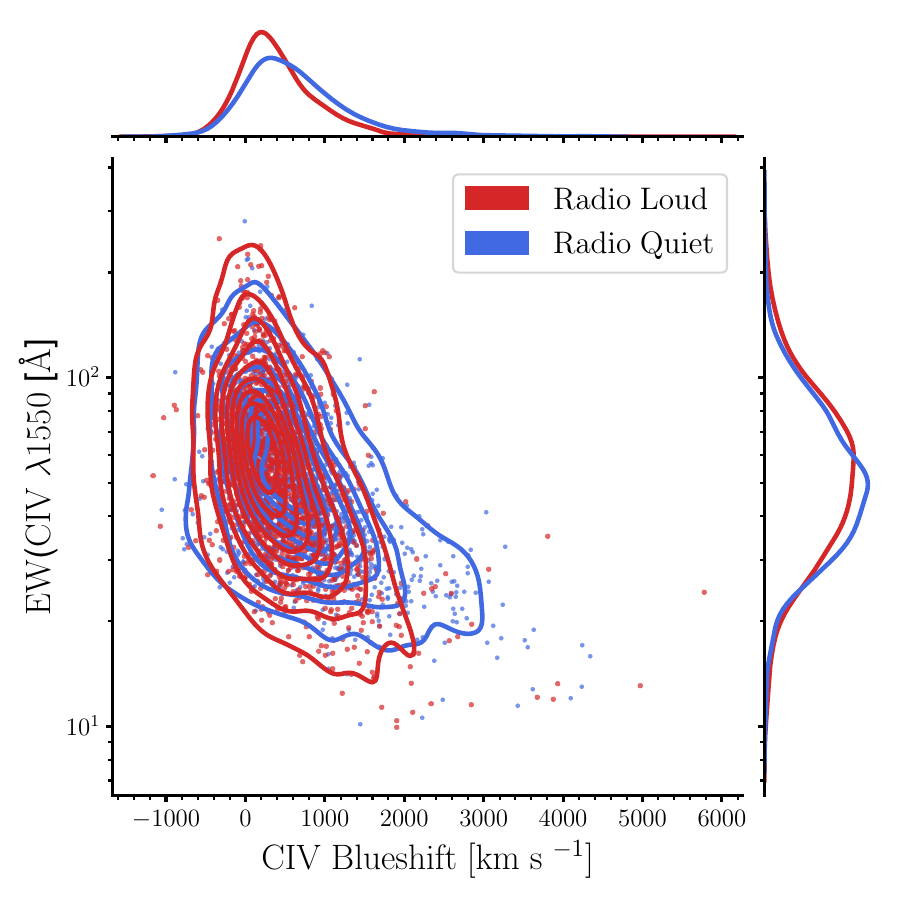}
    \caption{The $\textrm{C}\,\textsc{iv}$ blueshift and $\textrm{EW}_{\textrm{$\textrm{C}\,\textsc{iv}$}}$ distributions for radio loud (red points/contours) and radio quiet (blue points/contours) quasars, as defined by the GMM. Similarly to Fig. \ref{fig:pre-match}, the left plot is for the entire sample, and the right shows the distribution of sources after they are matched in $L_{3000}$ and $\textrm{FWHM}_{\textrm{$\textrm{Mg}\,\textsc{ii}$}}$. There are clearly differences in both distributions, even after effectively accounting for black hole mass and Eddington fraction.}
    \label{fig:oystersinmypocket}
\end{figure*}
In this section, we use the matched populations of RL and RQ quasars in various emission line spaces to investigate the differences between the outflows and accretion disc properties of these two populations, decoupled from effects of black hole mass and optical luminosity. \\

\begin{figure*}
    \centering
    \includegraphics[width=\columnwidth]{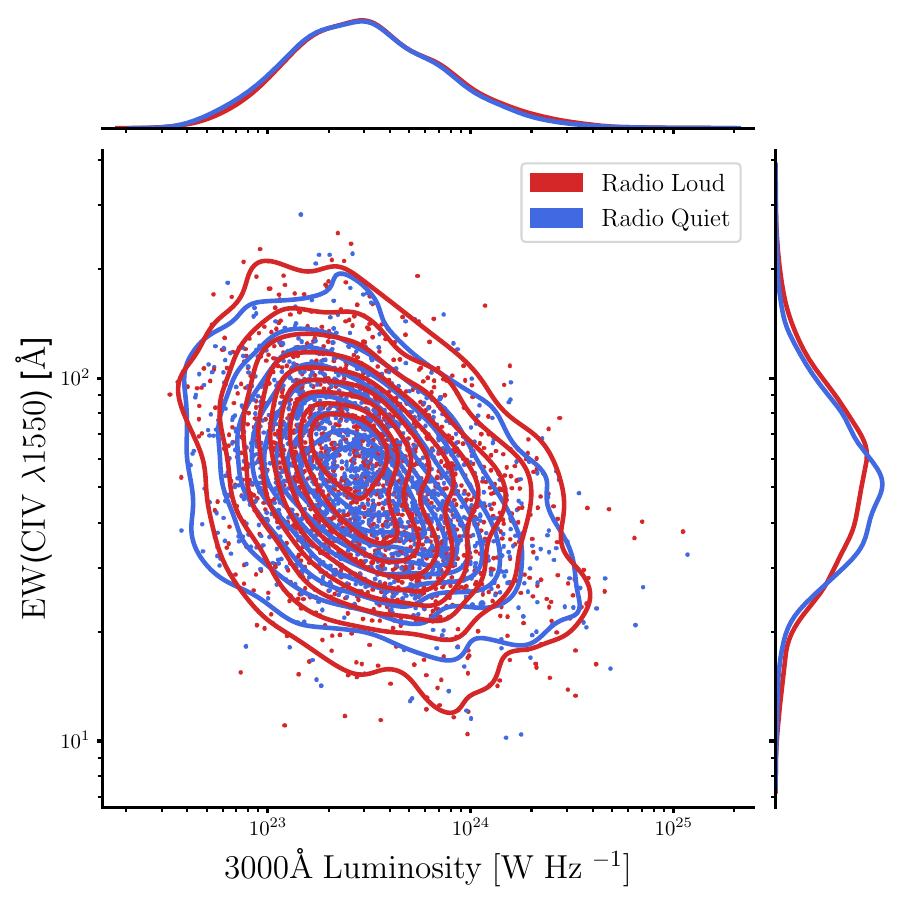}
    \includegraphics[width=\columnwidth]{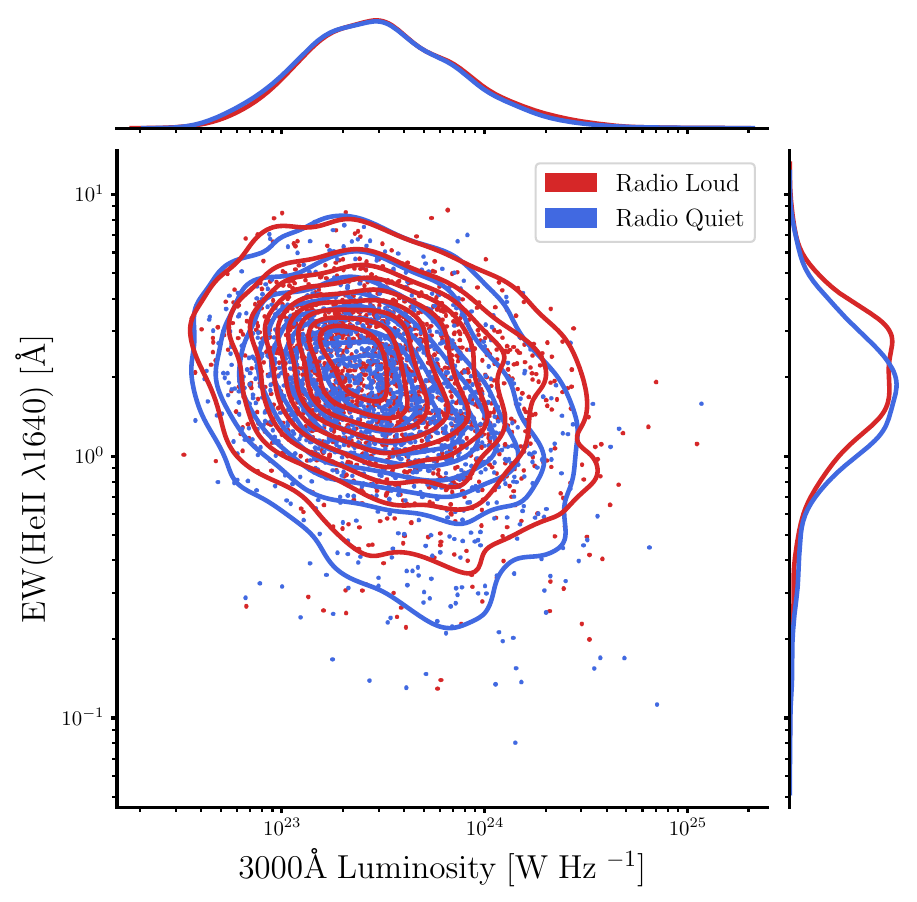}
    \caption{Distribution of radio loud (red points/contours) and radio quiet (blue points/contours) quasars in $L_{3000}$ and the EW of ultraviolet emission lines, $\textrm{C}\,\textsc{iv}$ (left) and $\textrm{He}\,\textsc{ii}$ (right), for populations  matched in $L_{3000}$ and FWHM$_{\textrm{$\textrm{Mg}\,\textsc{ii}$}}$. The RL and RQ populations here are defined by the GMM.}
    \label{fig:Beff}
\end{figure*}

In Fig. 7 of \cite{richards2011}, they compare the blueshift of the $\textrm{C}\,\textsc{iv}$ emission line to its EW, finding that there are no quasars with both a strong $\textrm{C}\,\textsc{iv}$ line and large blueshift. They also find, using 1.4 GHz radio data from FIRST \citep{becker}, that RL quasars do not occupy a separate part of the parameter space to RQ quasars, although they are more often found at the weaker blueshift part of the regime. A similar plot is reproduced in Fig. 5 of \cite{rankine2021} using LoTSS DR1, with the same conclusions. We now conduct the same exercise with the larger LoTSS DR2 sample in Fig. \ref{fig:oystersinmypocket}, both for the whole sample and then for the populations matched in $L_{3000}$ and $\textrm{FWHM}_{\textrm{$\textrm{Mg}\,\textsc{ii}$}}$. 

The overall \ion{C}{iv} emission properties are, unsurprisingly, similar to those found by \cite{richards2011} and \cite{rankine2020bal,rankine2021}: sources can either have strong $\textrm{C}\,\textsc{iv}$ emission or strong blueshifts, but not both. \footnote{As explained in section 4.3 of \cite{rankine2021}, many of the small number of objects in the lower-left region of Figure \ref{fig:oystersinmypocket} possess unreliable measurements and have deliberately been excluded from the analysis.} And whilst it is possible to find a RL source across the same range of $\textrm{C}\,\textsc{iv}$ emission space that one might find a RQ quasar, they are concentrated overall at lower blueshifts. Furthermore, even with the populations matched in $L_{\textrm{3000}} $ and $\textrm{FWHM}_{\textrm{$\textrm{Mg}\,\textsc{ii}$}}$, RL quasars have generally lower $\textrm{C}\,\textsc{iv}$ blueshifts than their RQ counterparts. The KS test $p$-values for the unmatched and matched $\textrm{C}\,\textsc{iv}$ blueshift distributions are $10^{-51}$ and $10^{-11}$ respectively.

On a source-by-source basis, effects such as inclination, and the impact of a constant bolometric correction and scatter in the relationships used to estimate black hole mass and Eddington fraction, act to confound any potential comparisons. However, given that these trends are statistically significant at a population level, the result is important: given the matching procedure, differences in the emission line properties of the populations must be driven by some process unconnected to black hole mass or Eddington fraction. There is also a small remaining difference in the $\textrm{EW}_{\textrm{$\textrm{C}\,\textsc{iv}$}}$ of the two populations, which, given that they are well-matched in ultraviolet luminosity, suggests that there is a slightly different Baldwin Effect in RL quasars as compared to RQ. When unmatched, the RL and RQ populations have a one-dimensional $\textrm{EW}_{\textrm{$\textrm{C}\,\textsc{iv}$}}$ KS-test $p$-value of $10^{-19}$, and which becomes $2\times10^{-4}$ when matched in $L_{3000}$ and $\textrm{Mg}\,\textsc{ii}$. In both cases, the RL and RQ populations are therefore statistically unlikely to be drawn from the same parent distribution. 

In Fig. \ref{fig:Beff} we plot EW versus luminosity for the two ultraviolet lines measured for our sample, $\textrm{He}\,\textsc{ii}$ and $\textrm{C}\,\textsc{iv}$. The RL and RQ sources display a Baldwin effect in both lines, which is to say there is a general decrease in EW with increasing luminosity, as shown previously in \cite{richards2011} for their sample of SDSS quasars with associated FIRST radio fluxes. However, for both emission lines, despite the populations being effectively matched in black hole mass and Eddington fraction, there is an offset between RL and RQ sources. For a given ultraviolet luminosity, a RL quasar will generally have a stronger $\textrm{He}\,\textsc{ii}$ and $\textrm{C}\,\textsc{iv}$ line than its RQ counterpart. 

\begin{figure}
    \centering
    \includegraphics[width=1\linewidth]{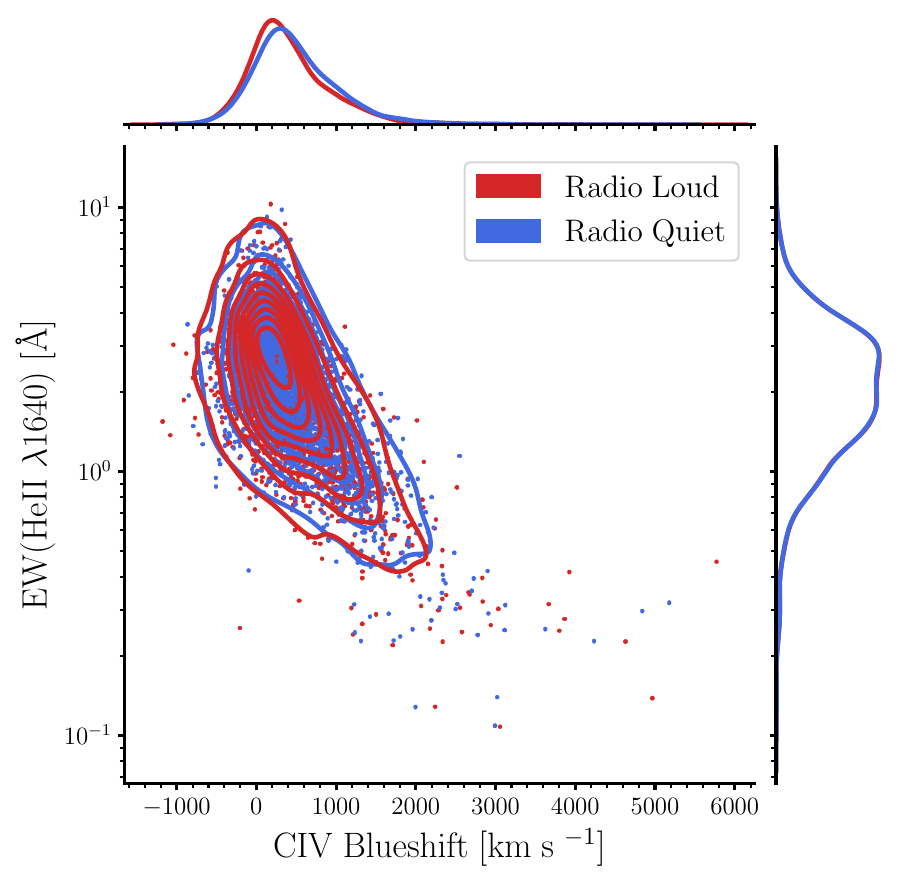}
    \caption{Distribution of $\textrm{C}\,\textsc{iv}$ blueshift and $\textrm{EW}_{\textrm{$\textrm{He}\,\textsc{ii}$}}$, for radio loud (red points/contours) and radio quiet (blue points/contours) quasars as defined by the GMM, after the populations have been matched in $L_{3000}$, $\textrm{FWHM}_{\textrm{$\textrm{Mg}\,\textsc{ii}$}}$ and $\textrm{EW}_{\textrm{$\textrm{He}\,\textsc{ii}$}}$. The $\textrm{EW}_{\textrm{$\textrm{He}\,\textsc{ii}$}}$ distribution is well-matched, and the radio loud and quiet populations show greater similarity in $\textrm{C}\,\textsc{iv}$ blueshift than in Fig. \ref{fig:oystersinmypocket}, but the radio loud sources are still shifted to lower blueshifts, overall. }
    \label{fig:heii match}
\end{figure}

To investigate the root cause of differences in $\textrm{C}\,\textsc{iv}$ blueshift between the RL and RQ populations, we carry out a three-dimensional match, now in $\textrm{FWHM}_{\textrm{Mg}\,\textsc{ii}}$, $L_{\textrm{3000}} $, and $\textrm{EW}_{\textrm{$\textrm{He}\,\textsc{ii}$}}$, with corresponding KS test $p$-values of 0.999 in each variable after matching. As such, the two samples are equivalent in black hole mass, Eddington fraction, and strength of ionising SED. We test the difference in $\textrm{C}\,\textsc{iv}$ blueshifts in Fig. \ref{fig:heii match}. The distributions of $\textrm{C}\,\textsc{iv}$ blueshift post-matching are more similar, but still RQ quasars have slightly higher blueshifts overall. The resulting KS-test $p$-value is $10^{-7}$, which suggests that, whilst the difference in strength of the ionising extreme ultraviolet and/or soft X-ray flux between RL and RQ sources is a large factor causing the difference in disc wind speeds, it cannot explain it in its entirety. 

\begin{figure*}
    	
    \includegraphics[width=\textwidth]{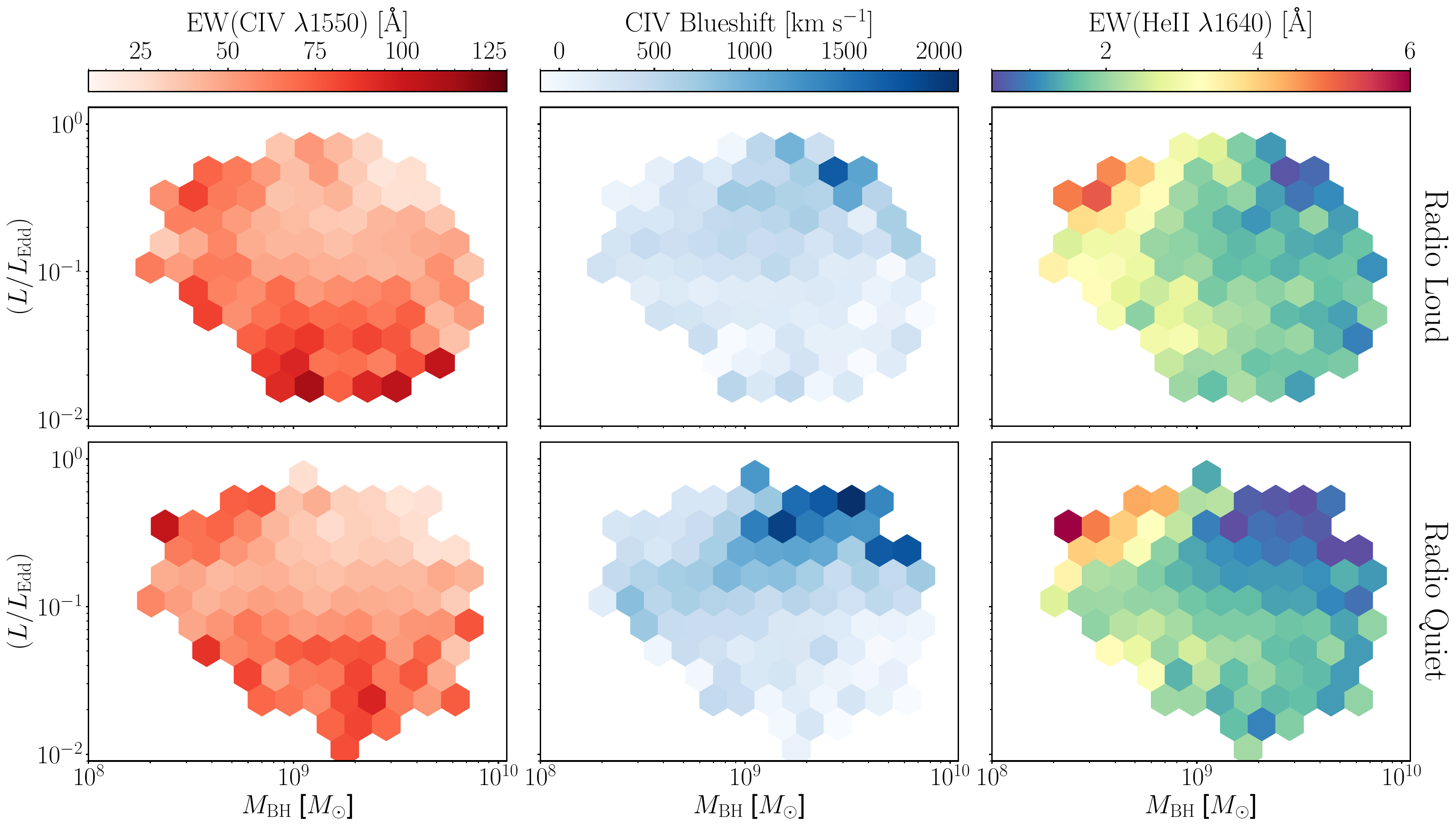}
    \caption{ Emission line properties for the matched samples of radio loud (top row) and radio quiet (bottom row) quasars in bins of black hole mass and Eddington fraction space. The colour scales show, from left to right, the median value of $\textrm{EW}_{\textrm{$\textrm{C}\,\textsc{iv}$}}$, $\textrm{C}\,\textsc{iv}$ blueshift, and $\textrm{EW}_{\textrm{$\textrm{He}\,\textsc{ii}$}}$. The known anti-correlation between $\textrm{C}\,\textsc{iv}$ blueshift and both $\textrm{He}\,\textsc{ii}$ and $\textrm{EW}_{\textrm{$\textrm{C}\,\textsc{iv}$}}$ holds for both radio loud and radio quiet sources. The trend of high blueshift objects at high black hole masses and Eddington fractions present in radio quiet sources is not present for radio loud quasars. The weakest $\textrm{EW}_{\textrm{$\textrm{He}\,\textsc{ii}$}}$ are seen in the same radio quiet sources with very large black hole masses and Eddington fractions that have high $\textrm{C}\,\textsc{iv}$ blueshifts.}
    \label{fig:tripletripletriple}
\end{figure*}

There are other very clear differences between the RL and RQ populations when matched in black hole mass and Eddington fraction, demonstrated in Fig. \ref{fig:tripletripletriple}. The figure shows how $\textrm{EW}_{\textrm{$\textrm{C}\,\textsc{iv}$}}$, $\textrm{C}\,\textsc{iv}$ blueshift and $\textrm{EW}_{\textrm{$\textrm{He}\,\textsc{ii}$}}$ vary across Eddington fraction and black hole mass space for both RL and RQ sources. The difference is again most prominent for $\textrm{C}\,\textsc{iv}$ blueshift, where the `wedge' shape of very high blueshift at the highest black hole masses and Eddington fractions found in \cite{temple2023} is significantly weaker for RL sources compared to their RQ counterparts, although there is also significantly weaker $\textrm{He}\,\textsc{ii}$ emission for the RQ sources in this region, too. There is a minor increase in the $\textrm{C}\,\textsc{iv}$ blueshift and a minor decrease in $\textrm{EW}_{\textrm{$\textrm{He}\,\textsc{ii}$}}$ for the most massive, highly accreting RLs only. 

As shown by \cite{temple2023}, there is an interesting change in $\textrm{EW}_{\textrm{$\textrm{He}\,\textsc{ii}$}}$ behaviour at $(\lambda_{\textrm{Edd}})\approx0.1$. In Fig. \ref{fig:tripletripletriple}, we show that the change occurs for both RL and RQ sources, albeit with the minimum value ($\textrm{EW}_{\textrm{$\textrm{He}\,\textsc{ii}$}}\sim1$\AA) only reached by RQ quasars, and a less significant appearance in RL quasars. In the wedge region defined in $\textrm{C}\,\textsc{iv}$ blueshift, RQ quasars show a clear deficit in $\textrm{He}\,\textsc{ii}$, inline with the idea that the sources with the weakest ionising SED are able to drive the strongest disc winds. There is a very similar distribution of $\textrm{EW}_{\textrm{$\textrm{C}\,\textsc{iv}$}}$ across the entire space for both populations, suggesting whatever is driving the EWs of $\sim100$\AA~  $\lambda_{\textrm{Edd}}\lesssim0.1$, versus the weakest emission, $\sim20$\AA, at $\lambda_{\textrm{Edd}}\approx0.1$ and $M_{\textrm{BH}}>10^{9}\textrm{M}_{\odot}$, is common across both RL and RQ sources. 
It is also evident that the correlations between the three emission line properties hold true for both sub-populations: that is, for the highest $\textrm{C}\,\textsc{iv}$ blueshifts, there is also the strongest $\textrm{He}\,\textsc{ii}$ emission line, and the weakest $\textrm{C}\,\textsc{iv}$ line, and vice versa. These constant correlations suggest there are enough similarities in the BLR of RL and RQ quasars that the coupling of the $\textrm{He}\,\textsc{ii}$ and $\textrm{C}\,\textsc{iv}$ emission lines is physically consistent. Equally, it is implying that the connection between disc winds and the underlying SED is constant in both RL and RQ sources, as these emission lines act as their tracers, perhaps meaning that differences in one are caused by a different distribution of the other. \footnote{See Fig. \ref{apfig:emissionLum} in Appendix \ref{Ap:diff cut} for the impact caused by a log$_{10}(R)$ and radio luminosity cut. In summary, the trends are unchanged for the populations defined using a constant radio-loudness cut; however, for the radio luminosity defined ones, there is more of a similarity between RL and RQ properties at high black hole masses and Eddington fractions. A more lenient cut in this part of the parameter space has likely resulted to contamination from objects with intermediate $\log_{10}(R)$ values.} 

\subsection{Accretion Properties of RL and RQ Sources}

\begin{figure}
	\includegraphics[width=\columnwidth]{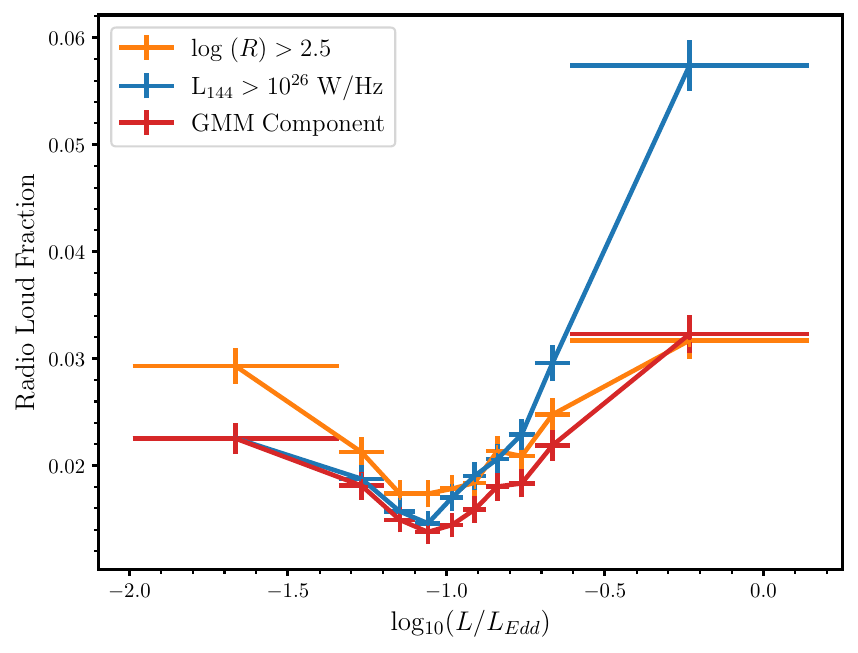}
    \caption{Radio loud fraction as a function of Eddington fraction, in bins of equal number of objects is shown for each of the three methods of radio loud classification: traditional radio loudness cut (orange), radio luminosity threshold (blue) and GMM component fit (red). All three methods show qualitatively the same trend: an initial decrease, followed by a minimum at 10 per cent of Eddington, then a further increase to a maximum at around the Eddington limit.}
    \label{fig:U Plot}
\end{figure}

As previously mentioned, it has been suggested by \cite{progastini} and \cite{petley} that the dominant form of outflow mechanism, and thus potentially also the source of radio emission, may be expected to vary with increasing accretion rate, transitioning from being jet-dominated to wind-dominated. In this framework, radio loudness decreases as a function of accretion rate. To test whether we observe such a trend, we show the RL fraction of objects as a function of $\lambda_{\textrm{Edd}}$, given by
\begin{equation}\label{eq:rlf}
    \textrm{RL fraction} = \frac{N_\textrm{RL}}{N_{\textrm{RL}}+N_{\textrm{RQ}}},
\end{equation}
where the fraction and numbers are defined within a given bin. In Fig. \ref{fig:U Plot}, we utilise the probability of each object being assigned to each GMM cluster, in order to mitigate the effects of uncertainties around borderline sources. As a result, the RL fraction is actually 
\begin{equation} \label{eq:prob rlf}
   \textrm{RL fraction} = \frac{\sum_i P_{\textrm{RL},i}}{N_{\textrm{RL}}+N_{\textrm{RQ}}}, 
\end{equation}
where the sum is over all sources in the relevant bin, and $P_{x,i}$ is the probability of a source $i$ belonging to sample $x$.
At Eddington fractions below $\lambda_{\textrm{Edd}}\sim0.1$, the fraction of RL sources decreases with increasing accretion rate. However, at $\lambda_{\textrm{Edd}}=0.1$ the RL fraction reaches a minimum -- interestingly, the same point at which \cite{temple2023} suggest there is a fundamental change in either BLR or accretion disc properties. Above $\lambda_{\textrm{Edd}}=0.1$, the RL fraction then increases to a maximum at $\lambda_{\textrm{Edd}} \approx 1$. The peak in RL fraction implies that powerful radio emission is most efficiently produced in quasars with Eddington fractions above $\lambda_{\textrm{Edd}}\approx0.1$. It has also been verified that these trends do not vary across the range of redshift in our sample.\footnote{These trends are consistent across the three methods, the main difference occurring at Eddington fractions of the order of unity, where a cut in radio luminosity shows a RL fraction of nearly 5.7$\pm$0.2 per cent, compared to the 3.2$\pm$0.2 per cent for the other two methods.} 

\begin{figure}

        \includegraphics[width=1\linewidth]{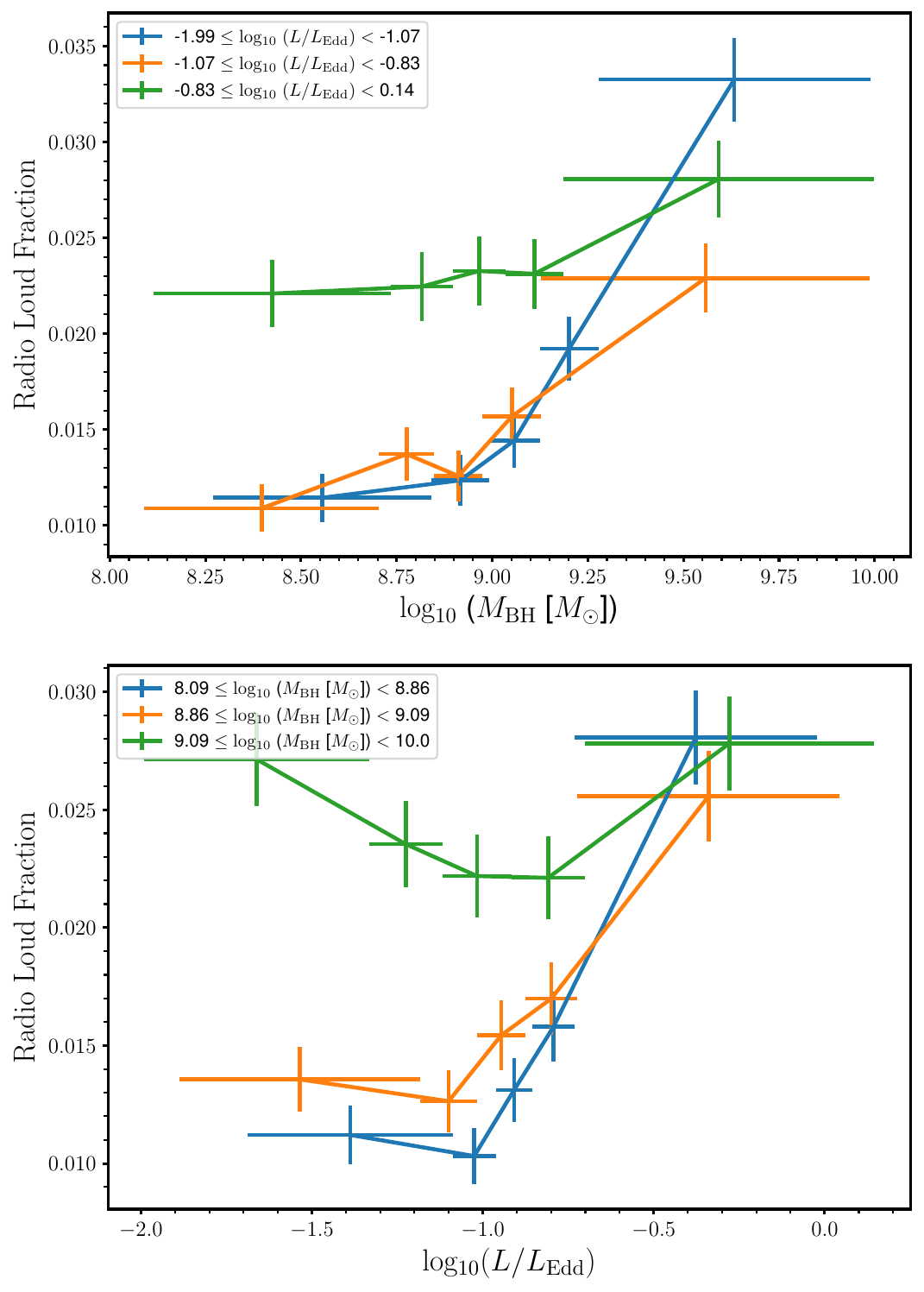}
    \caption{The fraction of objects categorised as radio loud by the GMM as a function of black hole mass (top) and Eddington fraction (bottom). Both are split into bins of equal number of objects, and each line represents varying Eddington fraction (top) or black hole mass (bottom), from lowest (blue) to highest (green). In both cases, low and intermediate bins of one property show a sharp upturn as the other property increases, whereas the highest bins are relatively high throughout.}
    \label{fig:2params}
\end{figure}
The investigation of the distribution of RL fraction is expanded in Fig. \ref{fig:2params} to examine its dependence on both $\lambda_{\textrm{Edd}}$ and black hole mass, which spans a range of $10^{8-10}\textrm{M}_{\odot}$ in our sample. There is a general increase in RL fraction with black hole mass, although the trend becomes shallower with increasing  $\lambda_{\textrm{Edd}}$. For \(\lambda_{\textrm{Edd}}<0.15 \), the range of RL fraction varies between only 2.2$\pm$0.2 and 2.8$\pm$0.2 per cent, compared to 1.1$\pm$0.1 and 3.3$\pm$0.2 per cent for \(\lambda_{\textrm{Edd}}>0.085 \). The RL fraction associated with the highest  $\lambda_{\textrm{Edd}}$ bin is significantly greater than for the other bins for black hole masses $M_\textrm{BH} \lesssim 10^{9.45}\textrm{M}_{\odot}$, above which it is overtaken by the lowest  $\lambda_{\textrm{Edd}}$ bin.

In terms of \(\lambda_{\textrm{Edd}}\), there is again a clear distinction between objects with the highest and lowest black hole masses. It is clear that above $\lambda_{\textrm{Edd}}< 0.1$ in Fig. \ref{fig:U Plot}, the RL fraction is dominated by the highest mass objects. In fact, in the highest black hole mass bin the RL fraction is $\sim2.5$ times higher than in those with intermediate mass black holes ($M_\textrm{BH} \lesssim 10^{9}\textrm{M}_{\odot}$) when $\lambda_{\textrm{Edd}}< 0.1$, at values of 2.7$\pm$0.2 per cent versus 1.1$\pm$0.1 per cent respectively. 

When $\lambda_{\textrm{Edd}}> 0.3$, RL fraction is comparable for all black hole masses, at $\approx$2.7 per cent. Generally, there is an increasing RL fraction with increasing $\lambda_{\textrm{Edd}}$, particularly when the black hole mass $M_\textrm{BH} \lesssim 10^{9}\textrm{M}_{\odot}$. At masses $M_\textrm{BH} \gtrsim 10^{9}\textrm{M}_{\odot}$, there is a high RL fraction across the entire  $\lambda_{\textrm{Edd}}$ range. 

\begin{figure*}
    \centering
    \includegraphics[width=1\textwidth]{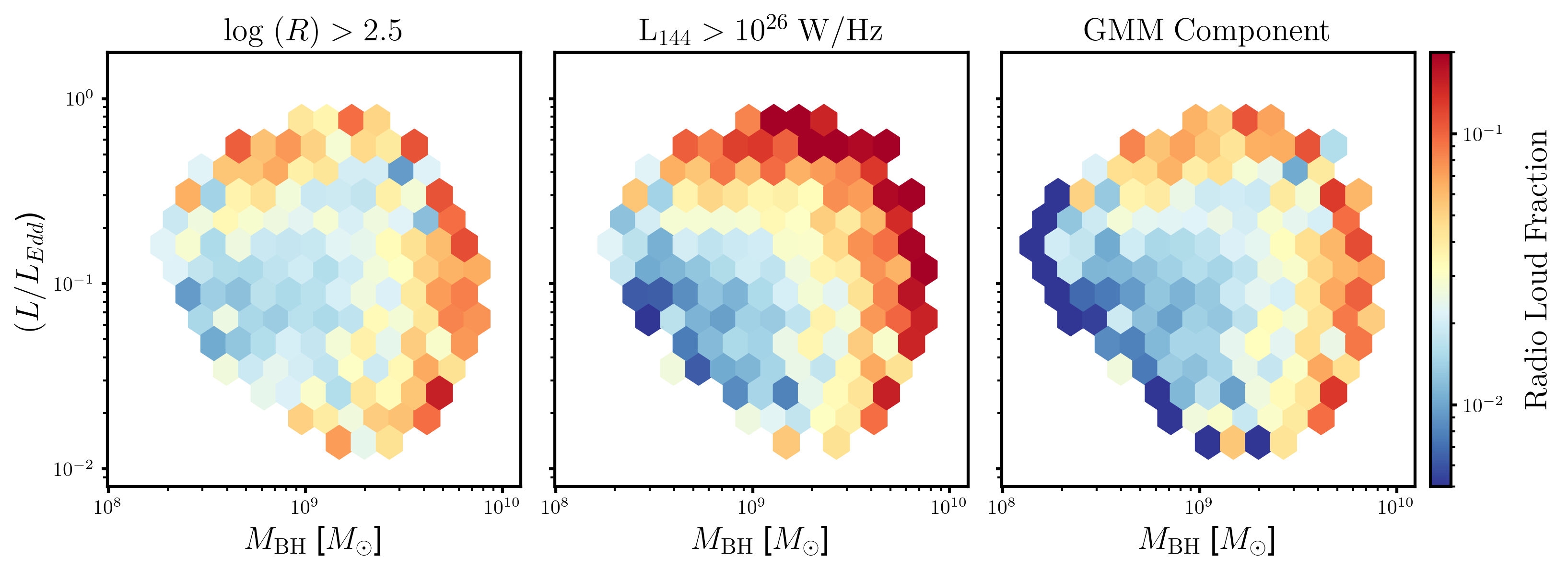}
    
    \caption{The distribution of average (mean) radio loud fraction in bins of black hole mass and Eddington fraction space, coloured on a log scale. The three panels show the results of the three different radio loud classification methods: classical radio loudness cut (left), GMM component assignment (middle) and radio luminosity threshold (right). The main difference here is the presence of many more radio loud sources with high black hole masses and Eddington fractions in the luminosity cut plot, driven by sources that are otherwise classified as radio quiet but have very high absolute radio luminosities.}
    \label{fig:double2}
\end{figure*}

Figure \ref{fig:double2} demonstrates in greater detail the origins of these trends, showing the 2D distribution of RL fraction across Eddington fraction and black hole mass for all three classification methods. There is a high RL fraction of $\sim$5-10 per cent at $\lambda_{\textrm{Edd}} \geq 0.3$, and again at  black hole masses of M$_{\textrm{BH}}>10^{9.4}\textrm{M}_{\odot}$. Intriguingly, when {\em both} of these criteria are met, the RL fraction is not always high; specifically, for the left and right hand panels, the extreme upper right-corner of the parameter space displays a typical RL fraction lower than the maximum value. The observed distribution can either be interpreted as the presence of two distinct regions of relatively high RL fraction, one at high  $\lambda_{\textrm{Edd}}$ and the other at high black hole mass, or alternatively that there is some suppression of jet formation in the quasars with the most massive, highly accreting black holes. In Section \ref{sec: disc-jet}, we discuss both possible interpretations. However, when we use a radio luminosity cut (right-hand panel of Fig \ref{fig:double2}) the reduction at high $\lambda_{\textrm{Edd}}$ and high black hole mass vanishes, and instead, there is a constant RL fraction of >8 per cent wherever the $\lambda_{\textrm{Edd}} > 0.3$ or  $M_{\textrm{BH}}>10^{9.4}\textrm{M}_{\odot}$ criteria are met. Interestingly, the region of reduced RL fraction in the GMM and radio loudness defined distributions corresponds remarkably well with the `wedge' of high $\textrm{C}\,\textsc{iv}$ blueshift, defined in \cite{temple2023} and Fig. \ref{fig:tripletripletriple}. 

Combined, Figs. \ref{fig:2params} and \ref{fig:double2} paint a compelling picture: RL fraction (and thus, inferred jet production) can be increased by a high accretion rate (as inferred from  $\lambda_{\textrm{Edd}}$) or a high black hole mass, and these two properties do not act in isolation, but are influenced by each other. 
\section{Discussion}
\label{sec:discuss}
Here, we consider the possible physical interpretations of our results and the implications of the different radio loud classification methods used in this paper. For the purposes of this discussion, we assume that the observational tracers of jets, disc winds, SED, and accretion rate (i.e. whether a source is radio loud, the blueshift of its $\textrm{C}\,\textsc{iv}$ emission line, the $\textrm{EW}_{\textrm{$\textrm{He}\,\textsc{ii}$}}$ and $\lambda_{\textrm{Edd}}$, respectively) act as reasonably accurate indicators of these properties, at least at a population level.

\subsection{Implications of Different Classification Methods}

For the majority of our results, the main conclusions are insensitive to the choice of RL classification method. The main difference between the samples produced by the three methods is that many more objects are classified as RL using the radio luminosity classification than the other two methods, particularly for sources that also display a large $L_{3000}$. 
 
The radio luminosity cut inherently biases the RL sample towards the most massive, highly accreting systems that, by definition, will be the most luminous at $L_{3000}$~(e.g. equations \ref{eq: black hole mass} and \ref{eq: Ledd}). These sources may also produce prodigious radio luminosity from other mechanisms as well as jets (see Section \ref{sec:radio in RL/RQ} for brief overview), meaning that luminous RQ quasars may contaminate the RL sample. However, the results from a constant luminosity cut do also tell us something interesting. In Fig. \ref{fig:rcut}, the upper envelope of the bulk distribution of RQ sources appears to flatten off at $L_{3000} \gtrsim10^{24}$W\,Hz$^{-1}$. At these high luminosities, a flatter cut should indeed be used to discriminate jetted/non-jetted objects, and many of the sources identified as RL by the luminosity cut are valid, and the other two methods are missing a substantial fraction. The exclusion of sources in this region could in part explain the `missing' wedge of sources seen in Fig. \ref{fig:double2} at high black hole mass/ Eddington fraction, that is filled in for the luminosity cut defined samples. Indeed, there are examples of traditionally designated `RQ', very massive sources (e.g. $M_{\textrm{BH}}\approx10^{9-10}\textrm{M}_{\odot}$) with powerful radio jets \citep{RQjets}. If the 2D RL fraction distribution from middle panel of Fig. \ref{fig:double2} is in fact most representative of the behaviour of AGN jet fraction, then there is no `missing' wedge, and the situation reverts to the simpler result: jet formation or prevalence is driven by high accretion rate and/or high black hole mass. There is also another interesting implication. For radio luminosity to no longer scale with optical luminosity, the source of radio luminosity (for these powerful RL quasars, relativistic jets) cannot continue to increase with increasing disc size/luminosity. That is, there is a maximum jet power able to be reached by quasars, which would be governed by an upper limit of jet launching efficiency, as suggested by \cite{fernandes2011evidence} -- and  some of the most massive and optically luminous quasars in our sample might be hitting it. \\

However, it is also entirely possible that the simple luminosity cut instead causes the sample to be biased to the most optically luminous systems, that inherently possess large radio luminosities. In Fig. \ref{fig:rcut}, it would appear that this cut does start to include some of sources at the upper edge of the RQ distribution, but if the cut were to be set any higher it would miss sources that are almost certainly {\sl bona fide} RL quasars at the lower end of the optical luminosity range. Inversely, the use of a traditional constant radio loudness cut may have the opposite effect: at $L_{3000} <10^{23}$~W\,Hz$^{-1}$, the radio emission threshold to meet even a strict cut (e.g. $\log_{10}(R)>2.5$) is such that it is feasible that objects with strong winds or high star formation rates could produce enough radio emission to be classified as RL. To be classified as RL, an object with a higher $L_{3000}$ (e.g. a quasar with a brighter disc) needs to have a radio luminosity up to two orders of magnitude brighter than one with a fainter disc. Therefore, at at high optical luminosities, the resulting sample from the $\log_{10}(R)$ cut could thus be biased towards only including the most luminous jets, risking losing those with more intermediate luminosities. Equivalently, there would be jetted objects missing from the sample at the highest optical luminosities, and RQ objects erroneously identified as RL at the lowest. By extension, that may mean the true RL fraction for the most massive, accreting objects may be suppressed, or artificially inflated for the smallest, least accreting sources. 
These potentially `missing' RL sources at the highest black hole masses and Eddington fractions could contributing to the creation of uncertainty in the upper right hand `wedge' portion of Fig. \ref{fig:tripletripletriple}. 

Given all the above, it is preferential to utilise a cut that relaxes the radio luminosity requirement at high optical luminosities, while still enforcing a high enough criteria at the lower end to ensure there is minimal contamination. Fortunately, the GMM classification naturally provides both of these things, as it defines two populations somewhere in-between the other two methods; furthermore, the approach is driven by the data and does not require a subjective choice of threshold. However, it should be noted that if a Bayesian GMM algorithm is used (i.e. one that self-determines the number of components required to describe the distribution), it routinely returns models with $\sim$30 components. The large number of components is indicative of the fact that the distribution of sources in radio and optical luminosity is \textit{not} a true dichotomy, but instead a continuous distribution. However, for our analysis, we seek a simple way to classify objects into RL and RQ, and the GMM fitting method would appear to be the most reliable, unbiased way of doing so. An additional strength of the GMM method is that it is possible to calculate the probability that a source is in each cluster, which allowed us to use these probabilities instead of a binary classification when calculating properties such as RL fraction (equation \ref{eq:prob rlf}). Such a statistical classification is undoubtedly preferable in situations when the RL/RQ nature of the source is ambiguous or the data point lies close to the boundary between populations. Returning to the concept of a `missing wedge' for the highest black hole masses and $\lambda_{\textrm{Edd}}$ (i.e. as described in Section \ref{sec:emission lines}), it's ubiquity, independent of the definition of RL objects, suggests the feature is an intrinsic property of the distribution.

\subsection{Physical Implications: Disc-Jet Coupling}
\label{sec: disc-jet}
The connection between accretion rate, black hole mass, and radio loudness in our sample of quasars is somewhat complex. It is clear that quasars are most able to launch jets when they have $M_\textrm{BH} > 10^9{\textrm{M}_{\odot}}$ or $\lambda_{\textrm{Edd}} > 0.3$, as shown in Figs. \ref{fig:2params} and \ref{fig:double2}. However, whilst there is significant uncertainty in the region with the highest black hole masses and Eddington fractions, there is tentative evidence to suggest that quasars with both a high black hole mass and high $\lambda_{\textrm{Edd}}$ are associated with a relatively lower RL fraction. We interpret the RL fraction as the relative prevalence of jetted quasars, or alternatively, the jet launching efficiency of quasars with a given set of properties. 

There are perhaps two equivalent ways of considering the result: that accretion rate and black hole mass are two mutually exclusive requirements that correspond to two different jetted populations, or that a high accretion rate and/or black hole mass contributes to jet launching, yet something acts to suppress jet formation for the most extreme sources. Focussing first on the former, we consider how this idea fits with previous studies. 

\subsubsection{Black Hole Mass Dependence}
There has been much attention paid to the connection between black hole mass and radio loudness. For example, \cite{mcldich} found that a black hole mass of \(M_{\textrm{BH}}> 10^9 \textrm{M}_{\odot}\) is needed to produce a powerful RL quasar, and both \cite{lacyradiobhmrel} and \cite{best2005mass} find a strong positive correlation between black hole mass and radio emission. \cite{McJarvisLumMassRel} also find that RL quasars are host to more massive black holes than RQ quasars. More recently, \cite{mighteeimogen} find in AGN generally, a black hole mass of \(M_{\textrm{BH}}> 10^{7.8} \textrm{M}_{\odot}\) is needed to launch and sustain a radio jet with mechanical power greater than the overall radiative output of the AGN, while \cite{Macfarlane2021} find a typical jet power that increases with the black hole mass in SDSS quasars. \cite{sabater} also find using cross-matched SDSS DR7 and LoTSS DR1 data that activity in the general radio AGN population (i.e. not limited to quasars) shows a strong positive correlation with black hole mass. The general consensus is that one might expect to find more jetted AGN at higher black hole masses, and our results support the hypothesis for quasars. 

However, \cite{bohan2} finds no connection between jet activity and black hole mass for the majority of their sample of SDSS DR16 quasars, apart from with the top 20 per cent most massive black holes. 
We too see this in our results, as there is no significant increase up to \(\textrm{M}_{\textrm{BH}} \approx 10^{9.1} \textrm{M}_{\odot}\), which is approximately the 80th mass percentile in our sample. However, the trend is not universal. The radio luminosity defined sample instead shows a general increase with black hole mass across the entire mass range. Furthermore, as seen in Fig. \ref{fig:2params}, the increase in RL fraction is less pronounced for sources in the highest $\lambda_{\textrm{Edd}}$ bin. The fact that BH mass trends change with Eddington fraction, and vice versa, highlights once again  the importance of considering them in combination when trying to understand the nature of quasar jet launching. To summarise, it is clear that regardless of the Eddington fraction, high black hole masses ($M_{\textrm{BH}} > 10^9\textrm{M}_{\odot}$) drive a high RL fraction, consistent with previous findings of a black hole mass dependence on jet launching. The inverse is also true: a high Eddington fraction ($\lambda_{\textrm{Edd}}\gtrsim0.1$) is associated with a high RL fraction, independent of black hole mass.

\begin{figure}
    \centering
    \includegraphics[width=1\linewidth]{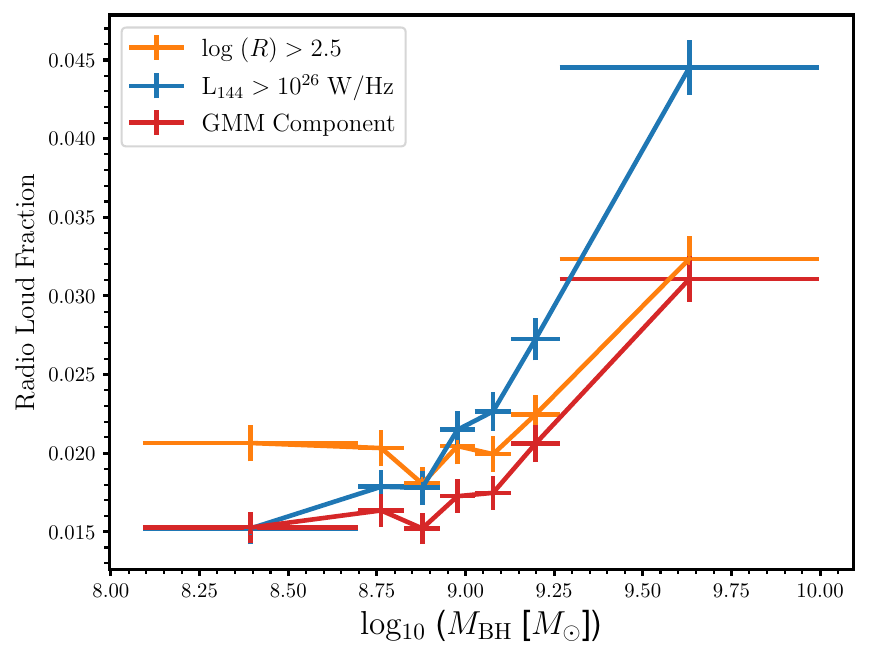}
    \caption{The fraction of quasars that are classified as radio loud, in bins of black hole mass with equal populations. The results are the same if the sample is binned using equal widths but different numbers. Results using the three different classifications methods are shown, with traditional radio loudness cut (orange), radio luminosity threshold (blue) and GMM component fit (red). The fraction is relatively constant for the radio loudness cut and the GMM until black hole masses of $\sim 10^{9.2}M_{\odot}$, whereas the luminosity cut threshold shows a steady increase across the entire range of masses. }
    \label{fig:black hole mass uplot}
\end{figure}
\subsubsection{Accretion Rate Dependence}
How does the literature expect accretion rate to link to the launching of jets in quasars? \cite{progastini} propose that the dominant source of radio emission in quasars changes with accretion rate, initially jet dominated at very low $\lambda_{\textrm{Edd}}$, until a transition to disc wind domination within systems approaching their Eddington limit. The result is supported by \cite{petley}, who suggest that radio emission across the range of Eddington fractions is generated by different processes, with disc winds at one end and jets at the other, assuming that disc winds produce detectable radio emission. Within the range of Eddington fractions that we observe in our sample, however, we do not see the same effect. We do observe an anti-correlation between radio properties and $\textrm{C}\,\textsc{iv}$ blueshift, shown in the differences between RL and RQ quasars in Fig. \ref{fig:oystersinmypocket}, and the co-location of high $\textrm{C}\,\textsc{iv}$ blueshift and reduced RL fraction in Figs. \ref{fig:tripletripletriple} and \ref{fig:double2}. However, the highest RL fractions are instead associated with the highest Eddington fractions ($\lambda_{\textrm{Edd}}\gtrsim0.1$), as seen in Figs. \ref{fig:U Plot}-\ref{fig:double2}. There are relatively fewer sources overall at high $\lambda_{\textrm{Edd}}$, however, there is a steeper drop-off in RQ sources than RL. That is to say, if a source has a high Eddington fraction, it is more prone to being RL than otherwise, or, by extension, that jet launching efficiency in this region of the parameter space is higher. 

Then, the question becomes why might a high Eddington fraction lead to a high jet launching fraction, or efficiency. It can be reasonably expected that accreted matter will carry magnetic flux \citep{rees1982ion}, which can lead to a collimation of matter from the disc and a subsequent massive ejection in the form of a jet, as described by the \cite{BP} (BP) formalism. Indeed, \cite{ferreira:insu-03326978} suggests that a BP-driven jet has its power directly controlled by the source's accretion rate, with higher accretion leading to greater jet power. Perhaps, then, the `accretion rate driven' component of the RL fraction that we are seeing in Fig. \ref{fig:double2} may be dominated by disc-launched, BP-like jets. Given that radio-loudness is a ratio and crudely encodes the efficiency of jet production relative to the accretion power, an increase of RL fraction with Eddington ratio may also require an increased jet {\em efficiency} in the same direction. In a BP-driven jet, this efficiency increase could be achieved through a dependence on the disc aspect ratio \citep{ferreira:insu-03326978}. Alternatively, within a BZ scenario, a correlation between jet efficiency and Eddington ratio can emerge due to more efficient accumulation of magnetic flux \citep[e.g.][]{ricarte} or an accretion-rate dependence of BH-spin (see also section~\ref{sec:spin}). 

\subsection{Physical Implications: Disc-Wind-Jet Connection}
\label{sec:DWJ}

The second possible interpretation of the results in Fig. \ref{fig:double2} is that there is something preventing the most massive, highly accreting quasars from consistently launching jets in an otherwise continuous distribution of high RL fraction at high Eddington fractions and black hole mass.

It is instructive to consider the differences in emission line properties as shown in Fig. \ref{fig:tripletripletriple}. The most striking difference between the matched samples of RQ quasars and RL quasars is clearly the absence of high $\textrm{C}\,\textsc{iv}$ blueshifts at high $\lambda_{\textrm{Edd}}$ and high black hole mass for RL quasars (Fig. \ref{fig:tripletripletriple}). \cite{temple2023} found that the most massive, highly accreting quasars typically display the largest $\textrm{C}\,\textsc{iv}$ blueshifts, which can be interpreted as showing the strongest disc winds. The finding is consistent with work by \cite{basklaorwinds} and \cite{progastini}, who suggest that a high $\lambda_{\textrm{Edd}}$ coupled with a high black hole mass is necessary for strong $\textrm{C}\,\textsc{iv}$ blueshifts if they are produced by disc winds. However, in our results, the need for high $\lambda_{\textrm{Edd}}$ and black hole mass to generate disc winds is only observed for sources that are RQ. One key question is therefore: does the presence of a jet inhibit disc winds, or do strong disc winds in quasars quench jet launching? 

The existence of an anti-correlation between jets and disc winds is something that is well documented, both in quasars and in other accreting systems. \cite{windsvsjets} find that, in microquasars, winds in the accretion disc can suppress jet activity by carrying away sufficient matter to halt the flow of mass into the jet. Related behaviour can be seen in quasars at the population level: Fig. 3 of \cite{rankine2021}, where it shows that there is a decrease in RL fraction with increasing $\textrm{C}\,\textsc{iv}$ blueshift, a results supported by Fig. 5 of \cite{rankine2021}, \cite{richards2011} and Fig. \ref{fig:oystersinmypocket} in this paper. All of these works point to a systemically lower $\textrm{C}\,\textsc{iv}$ blueshift in RL quasars compared to their RQ counterparts. Further suggestions of a wind-jet anti-correlation or bimodality have been made based on broad absorption line quasar properties \citep{morabito2019,petley2022} and X-ray spectroscopy of radio-loud Seyfert AGN \citep{Mehdipour2019}.

If the accretion disc winds are in fact a result of line driving, then we would expect to find a strong dependence on the ionising SED \citep[e.g.][]{murraywinds, proga2000, richards2011, progastini, rankine2021}, which would presumably manifest as an (anti-)correlation between the strength of $\textrm{C}\,\textsc{iv}$ blueshift and EW$_{\textrm{$\textrm{He}\,\textsc{ii}$}}$.  Indeed, in the region of Fig. \ref{fig:tripletripletriple} where we observe the highest $\textrm{C}\,\textsc{iv}$ blueshifts in RQ sources, we also see the weakest EW$_{\textrm{$\textrm{He}\,\textsc{ii}$}}$. Low EW$_{\textrm{$\textrm{He}\,\textsc{ii}$}}$ could be indicative of a softer SED, that does not overionise the gas in the disc and instead allows a line-driven wind to occur. The anti-correlation between $\textrm{C}\,\textsc{iv}$ blueshifts and EW$_{\textrm{$\textrm{He}\,\textsc{ii}$}}$ is not present in the RL sources, which show a comparable EW$_{\textrm{$\textrm{He}\,\textsc{ii}$}}$ across all $\lambda_{\textrm{Edd}}$ for black hole masses $M_{\textrm{BH}} >10^9\textrm{M}_{\odot}$. 

The increased ionization from the harder SED in RL sources likely precludes the formation of a disc wind, although there are also RQ sources with low $\textrm{C}\,\textsc{iv}$ blueshift, so a lack of wind does not itself guarantee efficient jet launching. In Fig. \ref{fig:Beff}, we show that for a given $L_{3000}$, a RL source will generally have a larger EW$_{\textrm{$\textrm{He}\,\textsc{ii}$}}$. As the RL and RQ quasars are matched in $L_{3000}$ and FWHM${_\textrm{$\textrm{Mg}\,\textsc{ii}$}}$, and thus effectively black hole mass and Eddington fraction, we find that two otherwise equivalent quasars will have different SEDs, depending on whether or not a radio jet is present, with RL sources possessing a 'harder' ionising spectrum. In Section \ref{sec:spin}, we discuss a possible cause. The results is consistent with the findings of \cite{rankine2021}, who found that strong EW$_{\textrm{$\textrm{He}\,\textsc{ii}$}}$ is a necessary but not sufficient condition for a jet to be present in their quasar sample, and that the presence of jets and quasar SED shape are likely to be correlated. 
\subsection{Black Hole Spin} \label{sec:spin}
In the well-known spin paradigm, a key determining factor of whether or not a quasar will produce a jet is the spin of its central black hole, with jets presumably associated with higher spin. A resulting general relativistic effect is that a quasar with a rapidly spinning black hole will also have an inner disc that extends closer to that black hole, which -- depending on the detailed accretion disc and corona physics -- could result in an increased EUV flux. Therefore, it is possible that the differences in EW$_{\textrm{$\textrm{He}\,\textsc{ii}$}}$ between high mass, high accretion rate RL and RQ quasars is a result of higher spin in those with jets \citep[see discussion by][]{richards2011,rankine2021}. However, whilst the presence of increased spin for sources that possess jets is appealing, there is no general increase in EW$_{\textrm{$\textrm{He}\,\textsc{ii}$}}$ for all RL quasars. The apparent decrease in the probability of jet-formation in sources with high $\lambda_{\textrm{Edd}}$ and high black-hole mass also requires further explanation in this scenario.

Furthermore, although the RL fraction of our sample is highest for $M_{\textrm{BH}}>10^9\textrm{M}_{\odot}$ and $\lambda_{\textrm{Edd}} >0.3$, it is at most $\sim10$ per cent. If rapid spin and a high black hole mass and/or a high accretion rate are necessary and sufficient conditions for efficient jet launching, then this behaviour would indicate that the spin values in the quasars are generally small, or otherwise the RL fraction would be much higher. However, such a scenario goes against population synthesis models such as \cite{volonteri2005distribution} and \cite{shapiro2005spin}, who both predict that the distributions of black hole spins in accreting systems, including quasars, should be skewed towards maximal. Therefore, either these conditions are insufficient, and instead there is another factor at play (e.g. magnetic field configuration, \citealt{Tchekjets}), or that our understanding of black hole spin which allows predictions such as those made by \cite{volonteri2005distribution} and \cite{shapiro2005spin} is incomplete. The reality of which of these scenarios is correct, if either, is beyond the scope of this paper, but is a key issue that still needs to be resolved.

\section{Conclusions}

We have used the rich multiwavelength data created by combining the LoTSS DR2 with the subset of SDSS DR17 that underwent spectral reconstructions in \cite{temple2023}, following the method of \cite{rankine2020bal}, to investigate the connection between disc winds, radio jets, Eddington fraction, and black hole mass in quasars between $1.5<z<2.5$. Our main conclusions are as follows:
\begin{enumerate}
    \item We have used a Gaussian Mixture Model to divide the distribution of sources into `Radio Loud' and `Radio Quiet' classifications, which is a statistically rigorous and easily reproducible methodology that allows for the incorporation of probabilities of that a source is radio loud/radio quiet. We have compared the results of this method of classification to that of a traditional radio loudness or radio luminosity cut, and have found that they are generally similar (Fig. \ref{fig:rcut}, Fig. \ref{fig:U Plot}, and Fig. \ref{fig:double2}).\\

    \item The fraction of sources inferred to be jetted varies non-linearly with Eddington fraction, increasing at low and high Eddington fractions, with a minimum at $\lambda_{\textrm{Edd}} \approx0.1$ and a characteristic `U' shape (Fig. \ref{fig:U Plot}). This is also the location at which \cite{temple2023} suggests a change in the physics of the broad line region, although it is unclear if these two results are directly connected. The radio loud fraction is also greatest for sources with the highest Eddington fractions, $\lambda_{\textrm{Edd}} \approx0.3$. \\
    
    \item For efficient jet launching, i.e. for a relatively large fraction of the sources to be presumed to possess a jet, quasars in our sample require either $M_\textrm{BH} > 10^9{\textrm{M}_{\odot}}$ or $\lambda_{\textrm{Edd}} > 0.3$ (Fig. \ref{fig:double2} and Fig. \ref{fig:2params}). This result is consistent with the idea that more massive black holes are able to generate more powerful radio emission, however it also hints towards an accretion-driven jet formation mechanism. We also find that the most massive, highly accreting quasars appear to less consistently form jets than those which are only massive, or highly accreting. \\
    
    \item In agreement with \cite{rankine2021}, we find that radio loud quasars generally have weaker $\textrm{C}\,\textsc{iv}$ blueshifts than radio quiet quasars, which we interpret as corresponding to weaker disc wind velocities. We now confirm that this result holds even when accounting for black hole mass, Eddington fraction (Fig. \ref{fig:oystersinmypocket} and Fig. \ref{fig:tripletripletriple}) and the strength of the ionising SED (Fig. \ref{fig:heii match}). The difference between populations is particularly clear for very massive, highly accreting sources ($\lambda_{\textrm{Edd}} >0.3$, $M_{\textrm{BH}}>10^9\textrm{M}_{\odot}$). We infer that there is an anti-correlation between the strength of accretion disc winds and the efficiency of jet launching, although weak wind signatures do not guarantee a high radio loud fraction. \\
    
    \item We find a different Baldwin effect in radio loud and radio quiet quasars, in both EW$_{\textrm{$\textrm{C}\,\textsc{iv}$}}$ and EW$_{\textrm{$\textrm{He}\,\textsc{ii}$}}$, although it is most pronounced in the latter. Thus, for a given $L_{3000}$, black hole mass, and Eddington fraction, a radio loud quasar will have a stronger $\textrm{C}\,\textsc{iv}$ and $\textrm{He}\,\textsc{ii}$ line than its RQ counterpart, and therefore a `harder' SED. Similar to the case for the $\textrm{C}\,\textsc{iv}$ blueshifts, these differences are particularly strong for sources with $\lambda_{\textrm{Edd}} >0.3$ and $M_{\textrm{BH}}>10^9\textrm{M}_{\odot}$. The differences in SED between the two populations may be a result of the radio loud sources having generally more rapidly spinning central black holes, as per the well-known spin paradigm. 

\end{enumerate}

\section*{Acknowledgements}

CLJ and JHM acknowledge funding from a Royal Society University Research Fellowship (URF\textbackslash R1\textbackslash221062). CLJ gratefully acknowledges support from the Balliol College Dervorguilla Scholarship. We thank Micah Bowles and Sophie Jewell for useful suggestions, and also thank Gordon Richards, Bohan Yue, Leah Morabito and Nicholas Choustikov for insightful discussions. We also extend our thanks to the anonymous reviewer for their feedback and suggestions. IHW and MJJ acknowledge support from the Oxford Hintze Centre for Astrophysical Surveys which is funded through generous support from the Hintze Family Charitable Foundation. MJJ acknowledges support from a UKRI Frontiers Research Grant [EP/X026639/1], which was selected by the European Research Council. MJT acknowledges funding from UKRI grant ST/X001075/1. ALR acknowledges support from a Leverhulme Early Career Fellowship. This work made use of {\tt astropy} \citep[][]{astropy2013,astropy2018,astropy2022}, {\tt seaborn} \citep{seaborn}, {\tt matplotlib} \citep{numpy}, {\tt scikit-learn} \citep{scikit-learn}, {\tt scipy} \citep{scipy}, and {\tt numpy} \citep{numpy}.   \\

LOFAR data products were provided by the LOFAR Surveys Key Science project (LSKSP; \url{https://LOFAR-surveys.org/}) and were derived from observations with the International LOFAR Telescope (ILT). LOFAR \citep{LOFAR} is the Low Frequency Array designed and constructed by ASTRON. It has observing, data processing, and data storage facilities in several countries, which are owned by various parties (each with their own funding sources), and which are collectively operated by the ILT foundation under a joint scientific policy. The efforts of the LSKSP have benefited from funding from the European Research Council, NOVA, NWO, CNRS-INSU, the SURF Co-operative, the UK Science and Technology Funding Council, and the Jülich Supercomputing Centre.\\

Funding for the Sloan Digital Sky Survey IV has been provided by the Alfred P. Sloan Foundation, the U.S. Department of Energy Office of Science, and the Participating Institutions. SDSS-IV acknowledges support and resources from the Center for High Performance Computing at the University of Utah. The SDSS website is \url{www.sdss.org}. 

SDSS-IV is managed by the Astrophysical Research Consortium for the Participating Institutions of the SDSS Collaboration including the Brazilian Participation Group, the Carnegie Institution for Science, Carnegie Mellon University, Harvard-Smithsonian Center for Astrophysics, the Chilean Participation Group, the French Participation Group, Instituto de Astrofísica de Canarias, The Johns Hopkins University, Kavli Institute for the Physics and Mathematics of the Universe (IPMU)/University of Tokyo, the Korean Participation Group, Lawrence Berkeley National Laboratory, Leibniz Institut für Astrophysik Potsdam (AIP), Max-Planck-Institut für Astronomie (MPIA Heidelberg), Max-Planck-Institut für Astrophysik (MPA Garching), Max-Planck-Institut für Extraterrestrische Physik (MPE), National Astronomical Observatories of China, New Mexico State University, New York University, University of Notre Dame, Observatário Nacional/ MCTI, The Ohio State University, Pennsylvania State University, Shanghai Astronomical Observatory, United Kingdom Participation Group, Universidad Nacional Autónoma de México, University of Arizona, University of Colorado Boulder, University of Oxford, University of Portsmouth, University of Utah, University of Virginia, University of Washington, University of Wisconsin, Vanderbilt University, and Yale University.
\section*{Data Availability}

The data underlying this article were accessed from the Sloan Digital Sky Survey (\url{https://www.sdss4.org/dr17/}) and the LOFAR Two-metre Sky Survey (\url{https://LOFAR-surveys.org/dr2_release.html}). The derived data generated in this research will be shared on reasonable request to the corresponding author.



\bibliographystyle{mnras}
\bibliography{mybib} 




\appendix

\section{Impact of Radio Loud Classification Scheme}
\label{Ap:diff cut}

This appendix shows versions of the plots in the body of the paper using the other two radio loud classification methods: the traditional radio loudness ratio cut (log($R$)$>2.5$) and the use of a radio luminosity threshold ($L_{\textrm{144}} > 10^{26} ~\textrm{W\,Hz$^{-1}$}$). 

\begin{figure}
    \centering
    \includegraphics[width=\columnwidth]{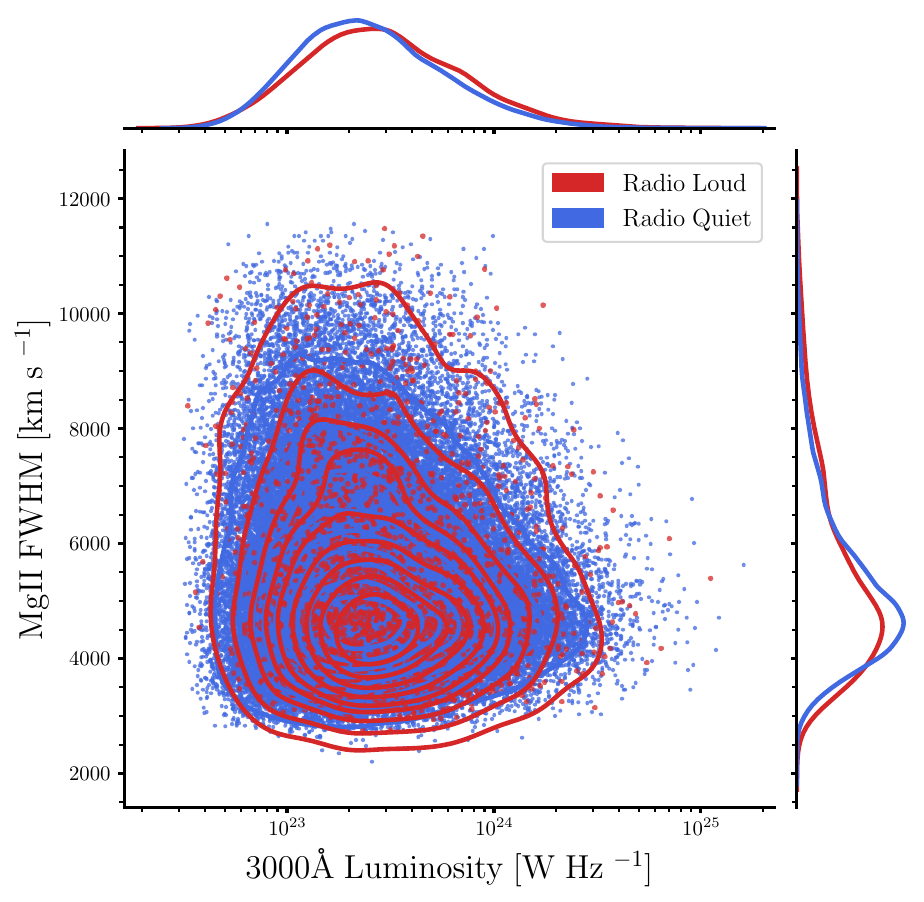}
    \includegraphics[width=\columnwidth]{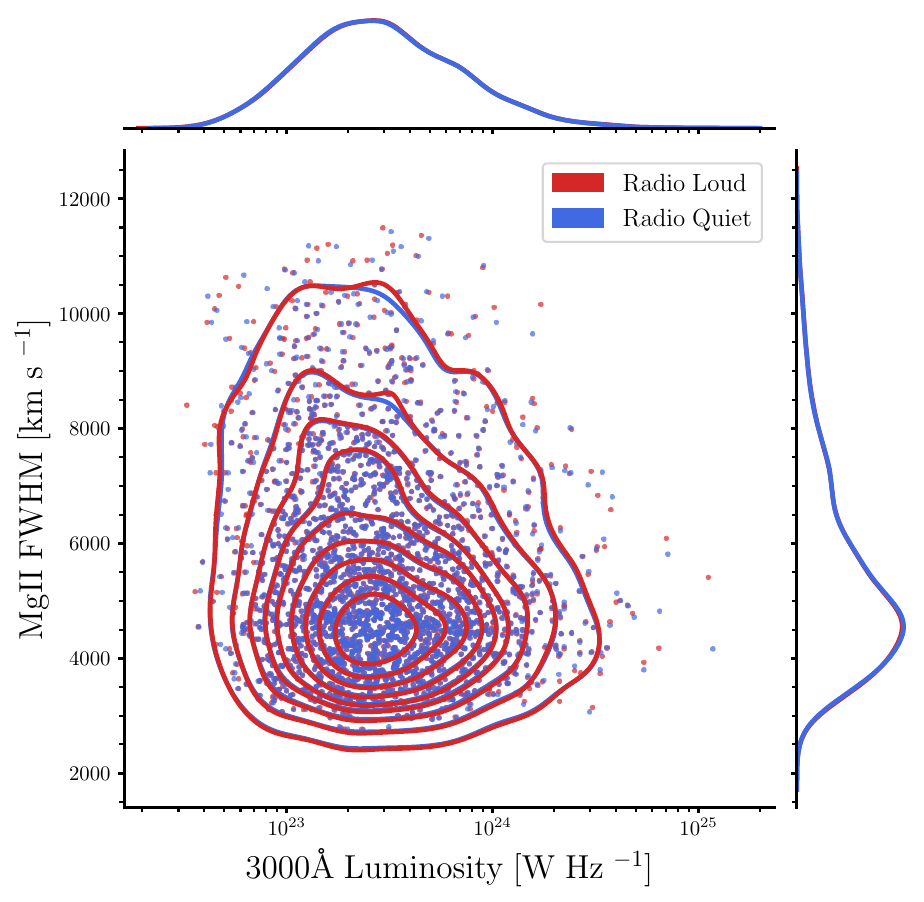}
      \caption{Distributions of radio loud (red points/contours) and radio quiet (blue points/contours) quasars in $L_{3000}$ and $\textrm{FWHM}_{\textrm{$\textrm{Mg}\,\textsc{ii}$}}$ space, for the entire sample (top) and then after NearestNeighbour matching in this space (bottom). The RL and RQ populations are as defined by the radio loudness ratio cut. As in Fig. \ref{fig:pre-match}, the matching procedure has resulting in indistinguishable distributions in both properties for RL and RQ sources.}
  \label{fig:matching rl cuts}

\end{figure}

\begin{figure}
     \centering
    \includegraphics[width=\columnwidth]{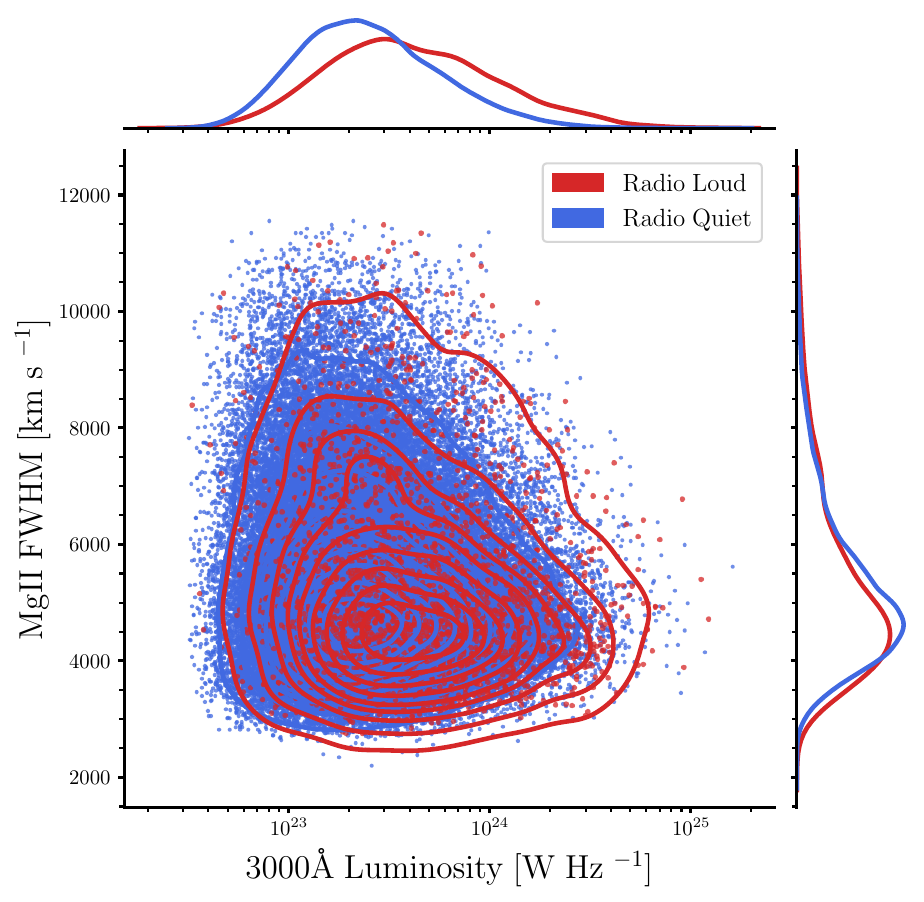}
    \includegraphics[width=\columnwidth]{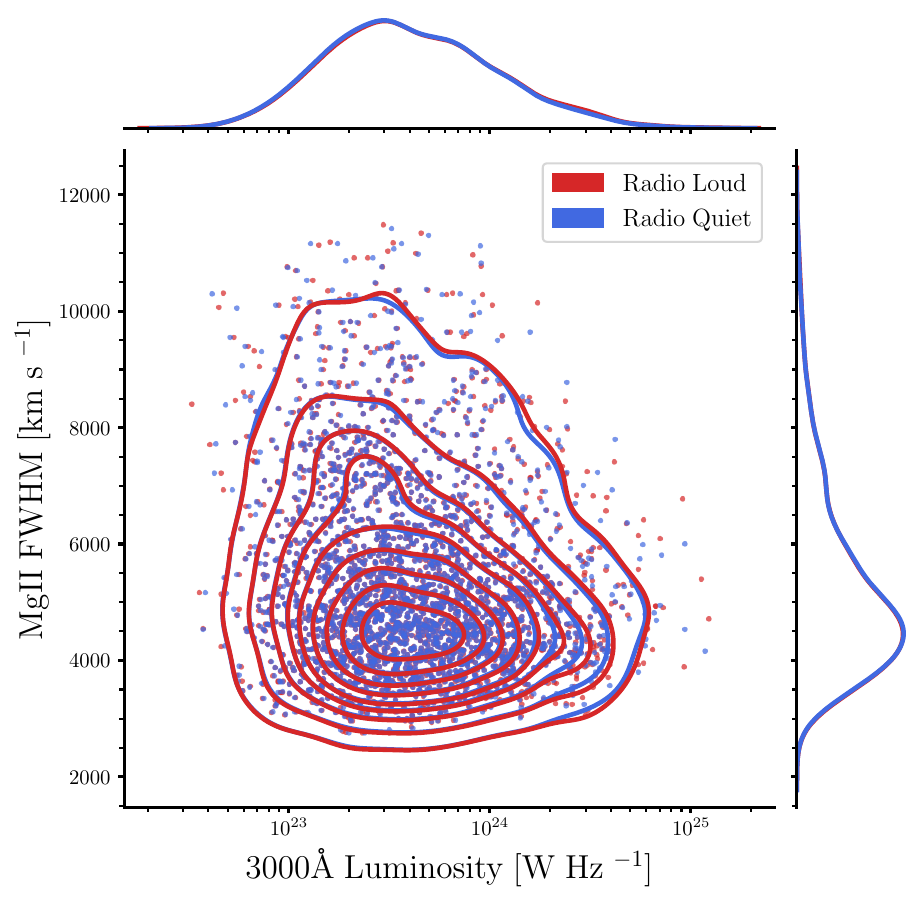} 
      \caption{Distributions of radio loud (red points/contours) and radio quiet (blue points/contours) quasars in $L_{3000}$ and $\textrm{FWHM}_{\textrm{$\textrm{Mg}\,\textsc{ii}$}}$ space, for the entire sample (top) and then after NearestNeighbour matching in this space (bottom). The RL and RQ populations are as defined by the radio luminosity cut ($L_{\textrm{144}} > 10^{26} ~\textrm{W\,Hz$^{-1}$}$). As in Fig. \ref{fig:pre-match}, the matching procedure has resulting in indistinguishable distributions in both properties for RL and RQ sources.}
  \label{fig:matching lum cuts}

\end{figure}

We perform the same NearestNeighbours matching procedure in $L_{3000}$ and FWHM$_{\textrm{$\textrm{Mg}\,\textsc{ii}$}}$ space as we did for the GMM-defined RL and RQ populations. The classical log($R$) cut has a resulting 2D KS-test $p$-value of 0.999 in the matched parameter space, and the L$_{144}$ cut has a $p$-value of 0.998. Both of these are thus statistically indistinguishable, and the distributions before and after matching are shown in Fig. \ref{fig:matching rl cuts} for the log($R$)$>2.5$ criteria, and in Fig. \ref{fig:matching lum cuts} for $L_{\textrm{144}} > 10^{26} ~\textrm{W\,Hz$^{-1}$}$. 

\begin{figure}
    \centering
    \includegraphics[width=\columnwidth]{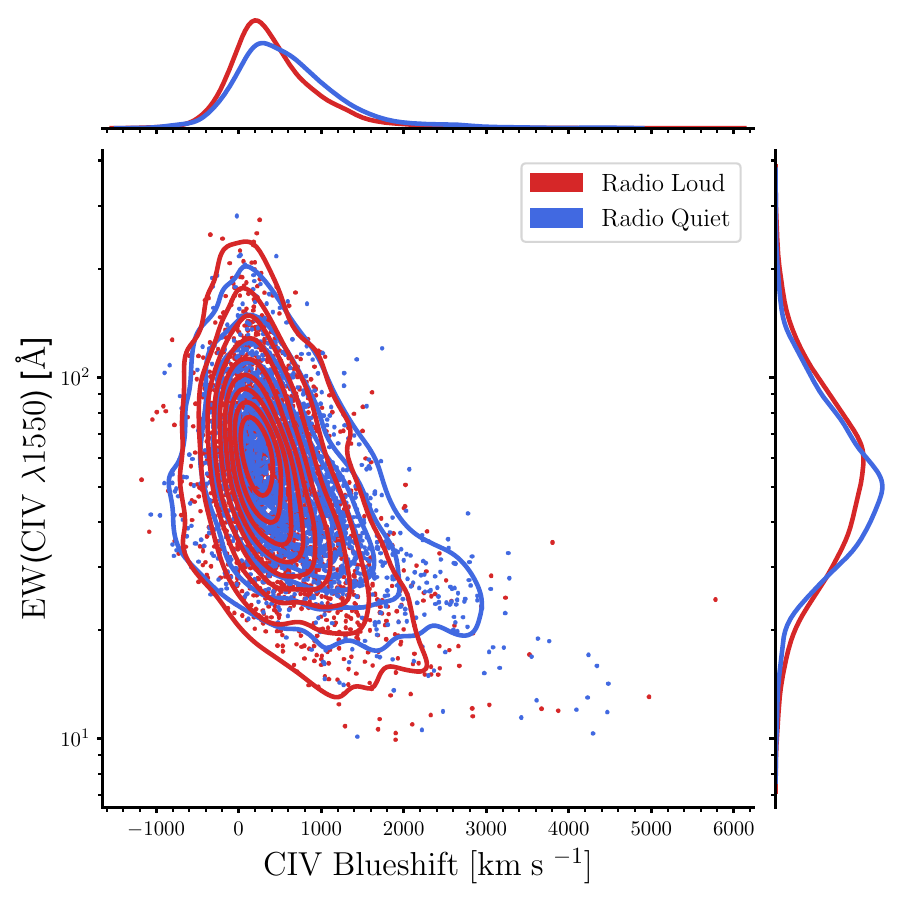}
    \includegraphics[width=\columnwidth]{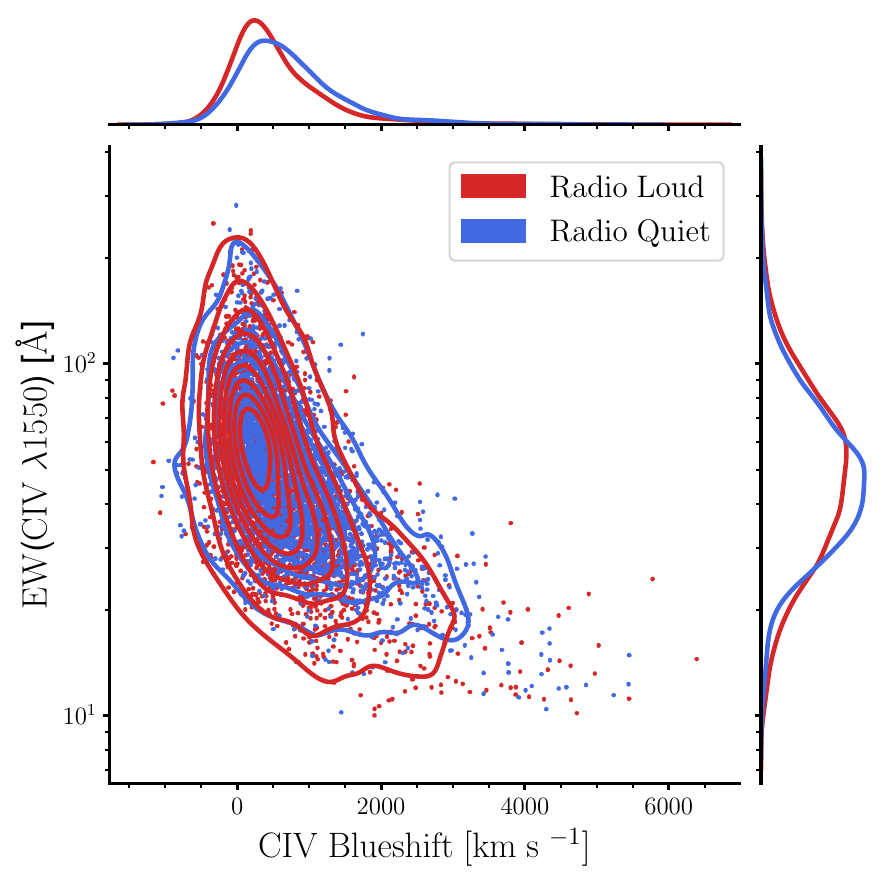}
    \caption{The $\textrm{C}\,\textsc{iv}$ blueshift and $\textrm{EW}_{\textrm{$\textrm{C}\,\textsc{iv}$}}$ distributions for radio loud (red points/contours) and radio quiet (blue points/contours) quasars, for the radio loudness ratio cut (log$_{10}(R)>2.5$, top) and then radio luminosity cut ($L_{\textrm{144}} > 10^{26} ~\textrm{W\,Hz$^{-1}$}$, right). The distributions show the populations after matching in $L_{3000}$ and $\textrm{FWHM}_{\textrm{$\textrm{Mg}\,\textsc{ii}$}}$. The trends found in Fig. \ref{fig:oystersinmypocket} are consistent across the classification methods. }
    \label{fig:civ space alt}
\end{figure} 

We show the distributions of $\textrm{C}\,\textsc{iv}$ blueshift and $\textrm{EW}_{\textrm{$\textrm{C}\,\textsc{iv}$}}$ for the matched populations as defined by the log($R$) and the $L_{\textrm{144}}$ cuts in Fig. \ref{fig:civ space alt}. It is clear that the difference in blueshift between the RL and RQ quasars is insensitive to the choice of radio loudness classification method: in all cases, RL quasars display weaker blueshifts than their RQ counterparts (and, presumably, weaker disc winds), even when accounting for differences in black hole mass and Eddington fraction. The distinction in $\textrm{EW}_{\textrm{$\textrm{C}\,\textsc{iv}$}}$ is less clear, however, with the resulting difference being greater for the luminosity cut defined samples, but less for the traditional radio loudness method. However, the resulting KS-test $p$-values are $2.30\times10^{-20}$ and $4.7\times10^{-4}$ respectively, and therefore the null hypothesis of the RL and RQ populations being drawn from the same distribution can be rejected. That means that, statistically, despite the RL and RQ $\textrm{EW}_{\textrm{$\textrm{C}\,\textsc{iv}$}}$ distributions appearing similar for all three classification methods, they are formally different. 
\begin{figure}
    \centering
    \includegraphics[width=\columnwidth]{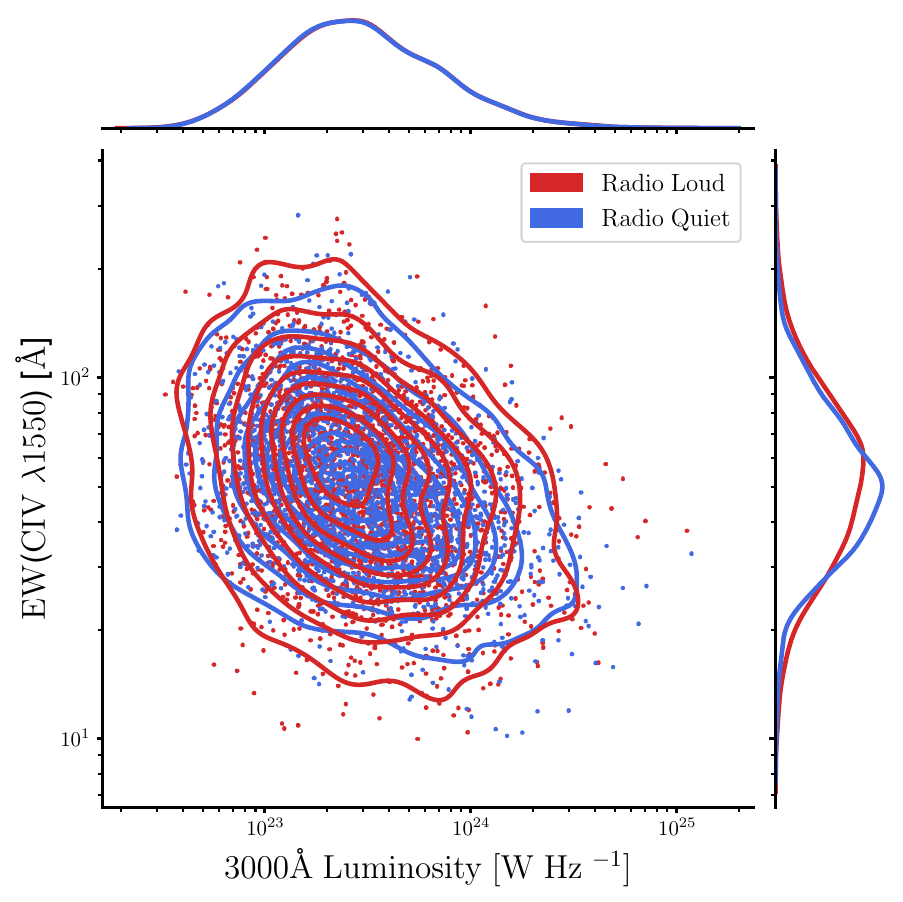}
    \includegraphics[width=\columnwidth]{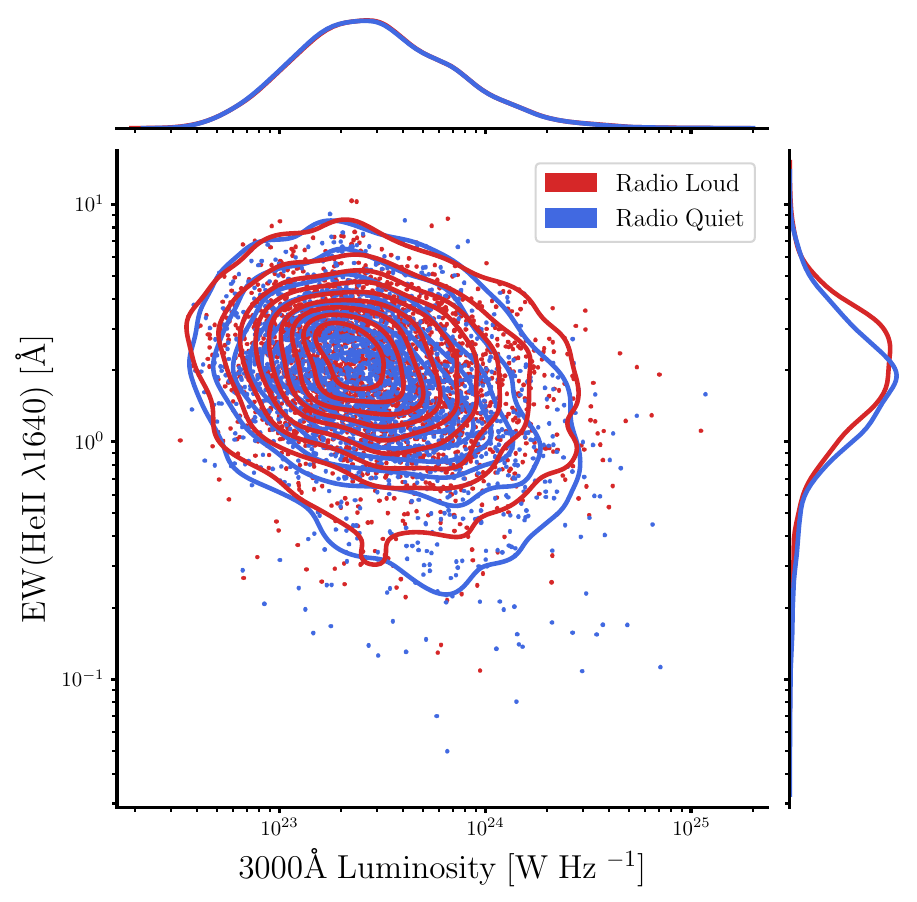}

  \caption{Ultraviolet emission line EWs for radio loud (red points/contours) and radio quiet (blue points/contours) quasars matched in $L_{3000}$ and $\textrm{FWHM}_{\textrm{$\textrm{Mg}\,\textsc{ii}$}}$, against $L_{3000}$. This is effectively a test of the Baldwin Effect. The plots on the top row show distribution of $\textrm{EW}_{\textrm{$\textrm{C}\,\textsc{iv}$}}$ against $L_{3000}$, and the bottom row shows the same, but for $\textrm{He}\,\textsc{ii}$. The populations in this plot are defined by the use of a log$_{10}(R$) cut. The trends found in Fig. \ref{fig:Beff} are also seen in this plot, and are therefore not dependant on the choice of classification method.}
  \label{fig:beff rl cuts}

\end{figure}

\begin{figure}
    \centering
    \includegraphics[width=\columnwidth]{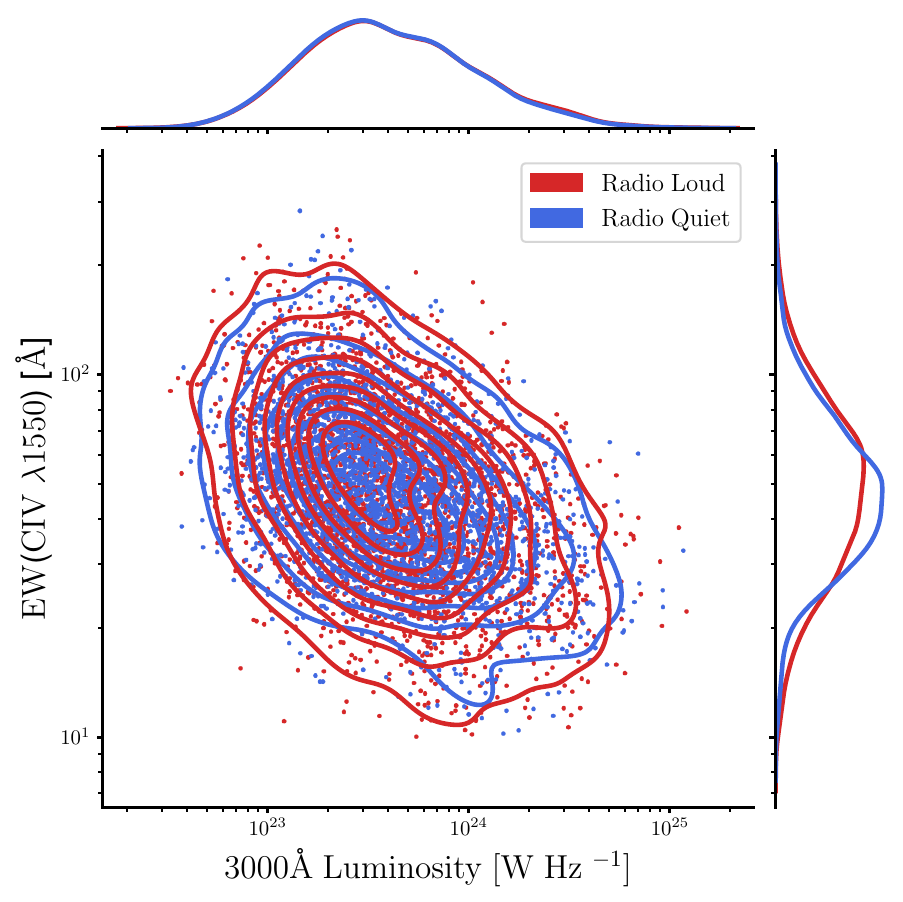}
    \includegraphics[width=\columnwidth]{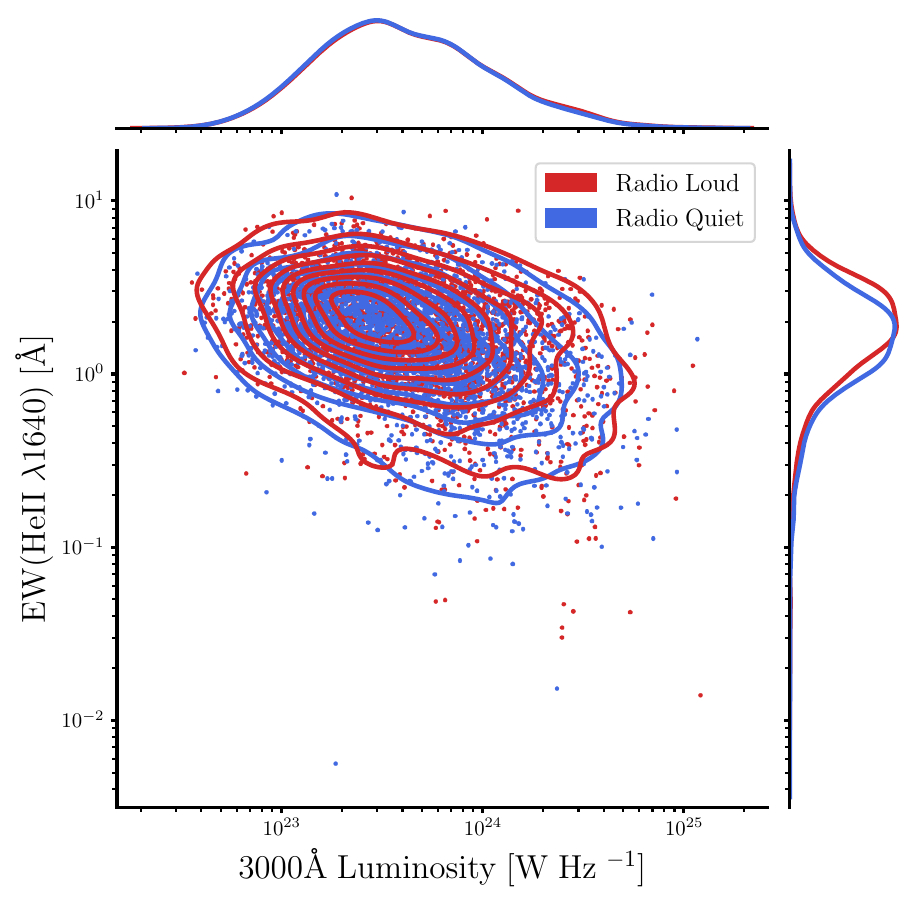}

  \caption{Ultraviolet emission line EWs for radio loud (red points/contours) and radio quiet (blue points/contours) quasars matched in $L_{3000}$ and $\textrm{FWHM}_{\textrm{$\textrm{Mg}\,\textsc{ii}$}}$, against $L_{3000}$. This is effectively a test of the Baldwin Effect. The plots on the top row show distribution of $\textrm{EW}_{\textrm{$\textrm{C}\,\textsc{iv}$}}$ against $L_{3000}$, and the bottom row shows the same, but for $\textrm{He}\,\textsc{ii}$. The populations in this plot are defined by the use of a $L_{\textrm{144}}$ cut. The trends found in Fig. \ref{fig:Beff} are also seen in this plot, and are therefore not dependant on the choice of classification method.}
  \label{fig:beff lum cuts}

\end{figure}
This has implications for the Baldwin Effect investigation, which is reproduced for the two alternate classification methods in Figs. \ref{fig:beff rl cuts} and \ref{fig:beff lum cuts}. Again, the conclusion that there is a slightly different Baldwin Effect in RL and RQ quasars is invariant under our different RL classification methods. 
\begin{figure*}
    \centering
    \includegraphics[width=1\textwidth]{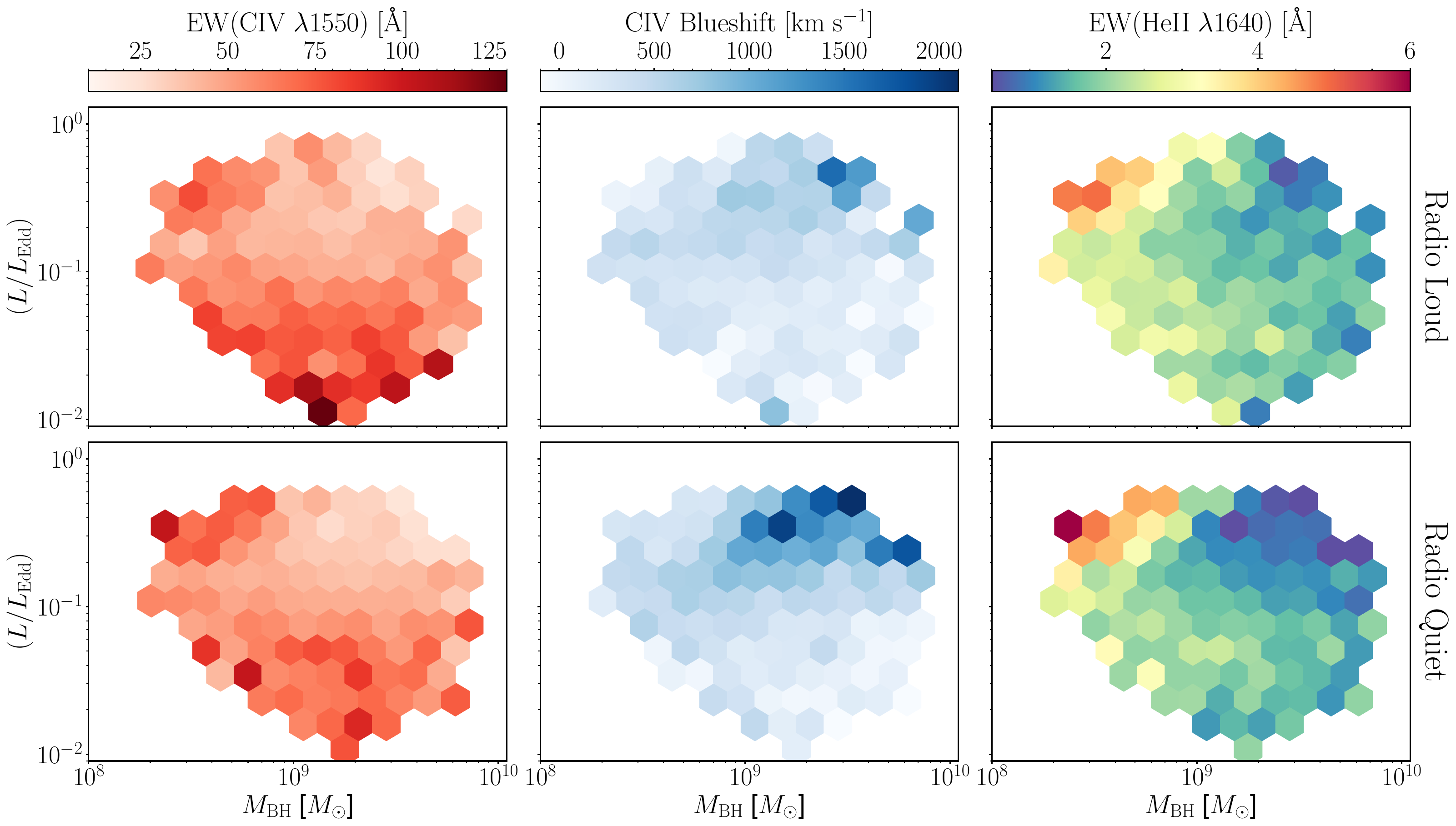}
    \includegraphics[width=1\textwidth]{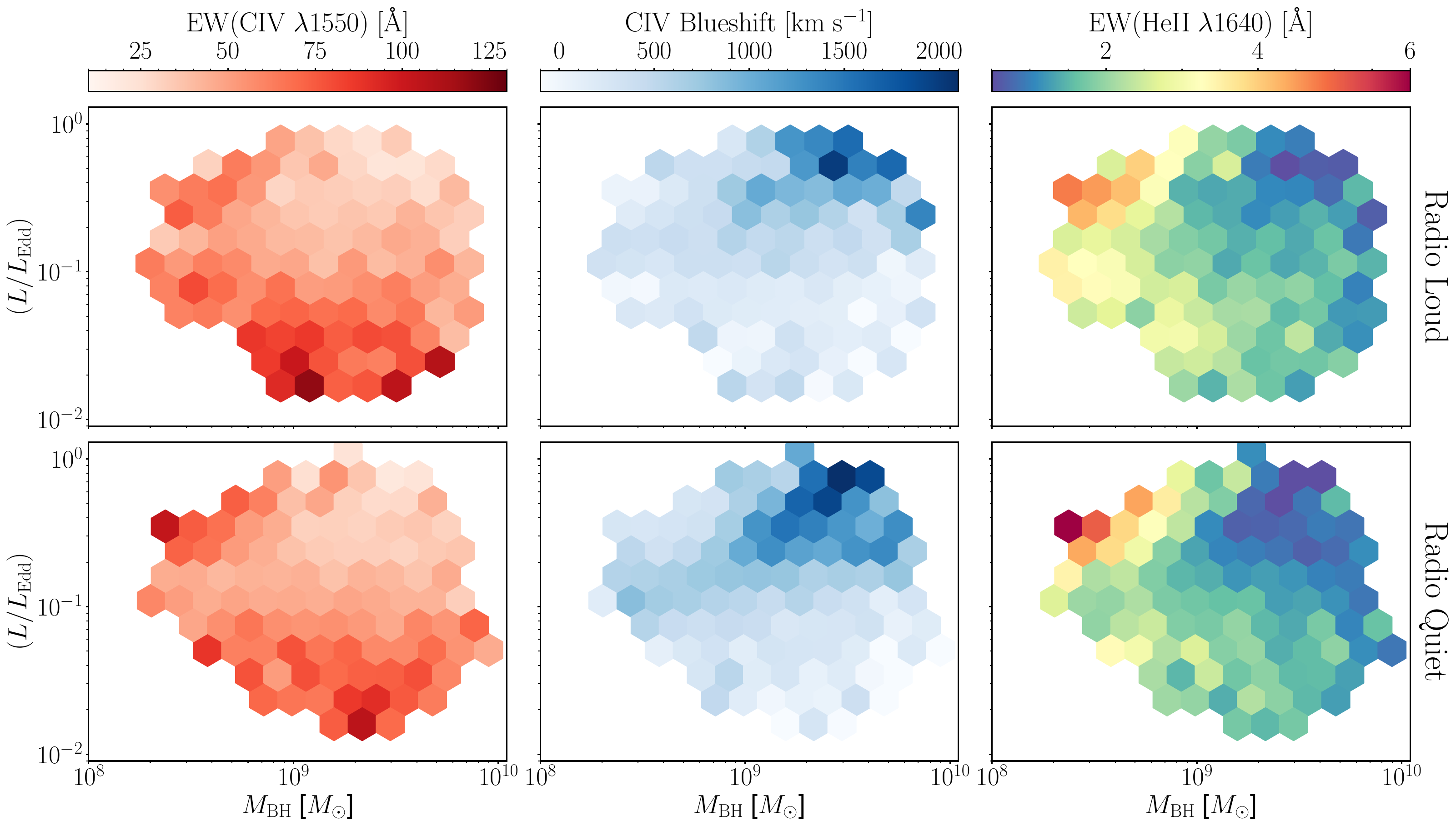}
    \caption{Ultraviolet emission line properties in bins of black hole mass and Eddington fraction space for the radio loudness (top) and radio luminosity (bottom)  defined samples, that have been matched in the equivalent of black hole mass and Eddington fraction. As in Fig. \ref{fig:tripletripletriple}, the top row of each panel shows the average value of each property for radio loud (top row) and radio quiet (bottom row) sources. The colour scales show, from left to right, the average (median) value of $\textrm{EW}_{\textrm{$\textrm{C}\,\textsc{iv}$}}$, $\textrm{C}\,\textsc{iv}$ blueshift, and $\textrm{EW}_{\textrm{$\textrm{C}\,\textsc{iv}$}}$. The trends seen in Fig. \ref{fig:tripletripletriple} are consistent across all three methods of radio loud classification.}
    \label{apfig:emissionLum}
\end{figure*}

We recreate the investigation of emission line properties in black hole mass and Eddington fraction space from Fig. \ref{fig:tripletripletriple} in Fig. \ref{apfig:emissionLum} for the radio loudness cut and radio luminosity cut defined populations respectively. The main results are the same across all three figures. In all cases, there is an anti-correlation between the strength of $\textrm{C}\,\textsc{iv}$ blueshift and the EW of the $\textrm{C}\,\textsc{iv}$ and $\textrm{He}\,\textsc{ii}$ lines, and RQ sources with very high black hole masses ($M_{\textrm{BH}} > 10^{9}\textrm{M}_{\odot}$) and Eddington fractions $\lambda_{\textrm{Edd}}> 0.1$ show very strong $\textrm{C}\,\textsc{iv}$ blueshifts, but the equivalent RL sources do not. However, in the $L_{144}$ defined method, the differences between trends in the RL and RQ sources are slightly less pronounced. It could be inferred that the presence of sources displaying `RQ-like' properties in the RL sample (i.e. strong $\textrm{C}\,\textsc{iv}$ blueshifts at high $M_{\textrm{BH}}$ and $\lambda_{\textrm{Edd}}$) may be a result of contamination from non-jetted sources that we have hypothesised maybe classified as RL as a result of using the method throughout the discussion of the cuts.

\begin{figure}
    \centering
    \includegraphics[width=1\linewidth]{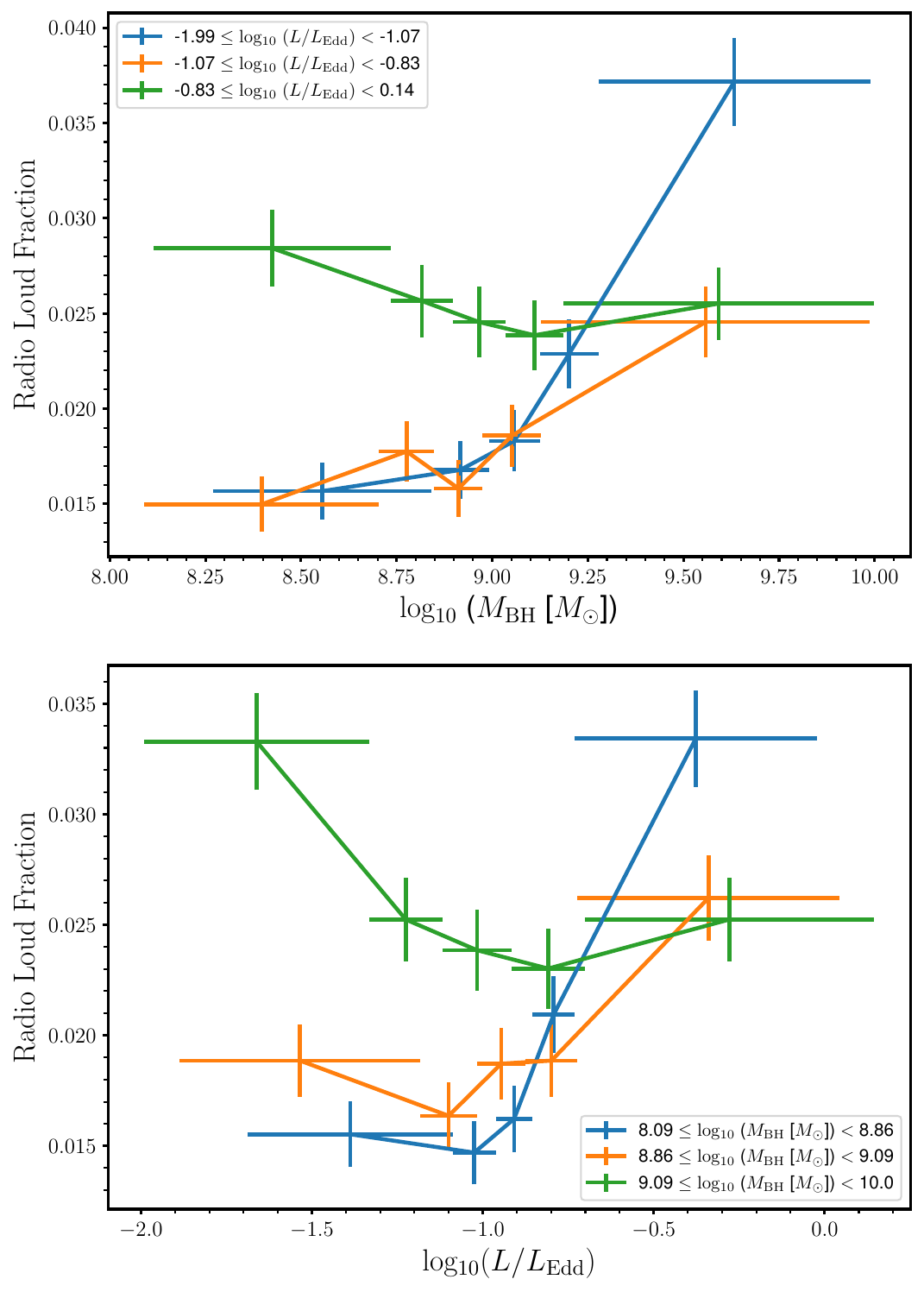}
    \caption{The distribution of radio loud fraction with increasing black hole mass (top panel) and Eddington fraction (bottom panel). As in Fig. \ref{fig:2params}, the coloured lines correspond to bins of Eddington fraction (top) and black hole mass (bottom), to show the cumulative effect of the two properties on the radio loud fraction. This version of the plot shows the results for RL and RQ populations defined by a classical radio loudness ratio criteria.}
    \label{apfig:multiurl}
\end{figure}
\begin{figure}
    \centering
    \includegraphics[width=1\linewidth]{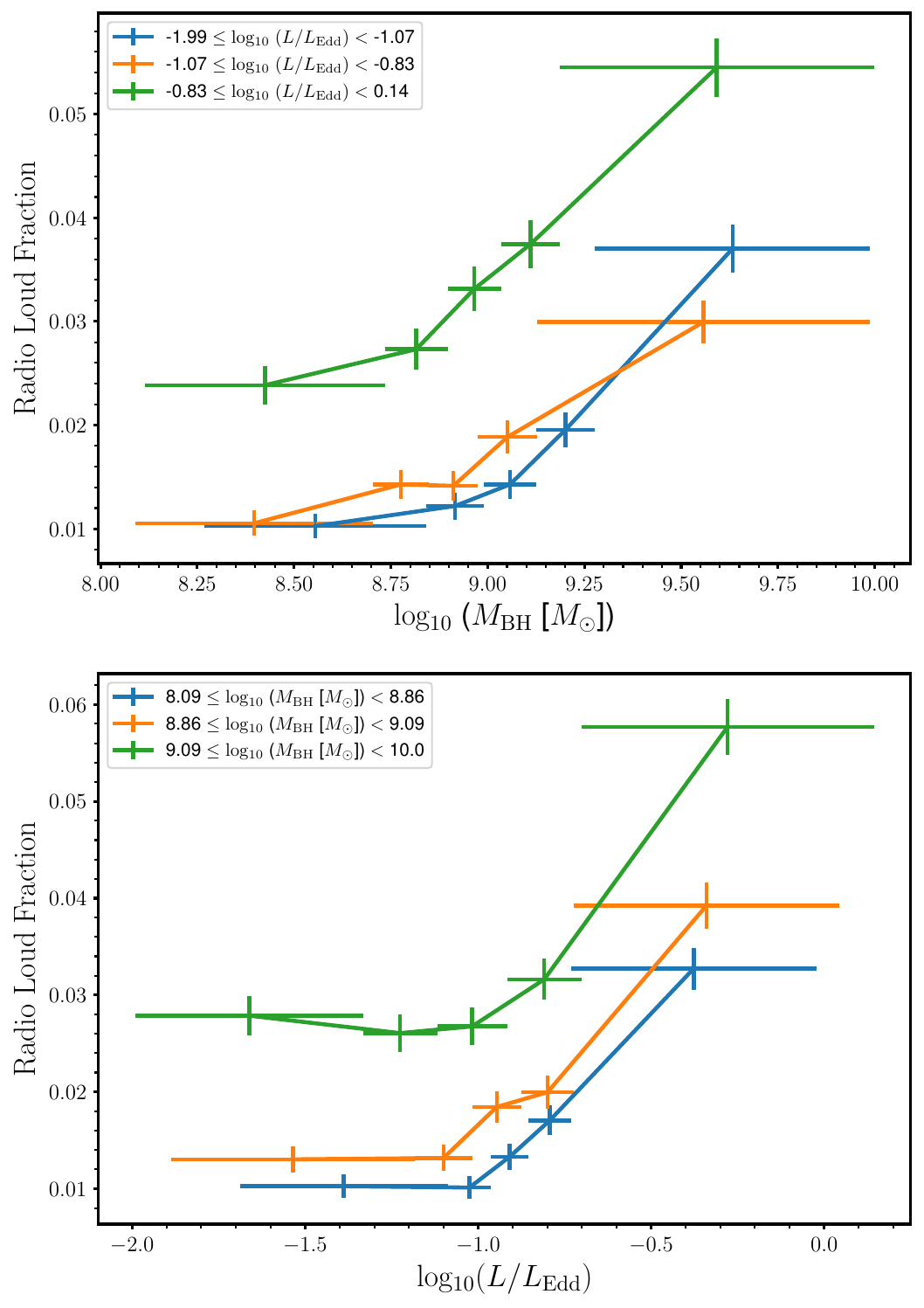}
    \caption{The distribution of radio loud fraction as a function of black hole mass and Eddington fraction, as in Figs. \ref{fig:2params} and \ref{apfig:multiurl}. This plot shows the results of the use of the radio luminosity cut.}
    \label{apfig:multiulum}
\end{figure}

Finally, in Figs. \ref{apfig:multiurl} and \ref{apfig:multiulum}, we reproduce Fig. \ref{fig:2params}. The results of the log$_{10}(R$) cut are generally consistent with the results produced by the GMM-defined populations, except for the highest black hole masses ($M_{\textrm{BH}} > 10^{9.36}\textrm{M}_{\odot}$) and highest Eddington fractions ($\lambda_{\textrm{Edd}}> 0.15$), where there is an even greater decrease in RL fraction whilst the other property is high. In the $L_{\textrm{144}} > 10^{26} ~\textrm{W\,Hz$^{-1}$}$ case, for all Eddington fraction and black hole mass bins, the fraction of radio loud sources increases with the two properties. In the latter case, this is a result of the general increase of $L_{144} $ with $L_{3000}$, meaning that as either black hole mass or Eddington fraction increases, more sources meet the radio luminosity threshold (see Eqs. \ref{eq: black hole mass} and \ref{eq: Ledd}). In the former case, it's the opposite: for the highest black hole mass and Eddington fraction objects, a source must have significantly higher radio emission to compete with the prodigious optical luminosity, causing a decrease in RL fraction.


\bsp	
\label{lastpage}
\end{document}